\documentclass[pre,twocolumn,superscriptaddress,showpacs]{revtex4}
\usepackage{graphicx,color}
\usepackage{amsmath}
\definecolor{b}{rgb}{0,0,1.0}
\definecolor{r}{rgb}{1,0,0}
\definecolor{g}{rgb}{0.2,1,0.1}

\begin{document}
\newcommand{\Froude}{{\rm Fr}}
\newcommand{\LANL}{Condensed Matter \& Thermal Physics and Center for Nonlinear Studies, Los Alamos National Lab, NM, 87545, USA}

\newcommand{\SZFKI}{Research Institute for Solid State Physics and Optics,
                     POB 49, H-1525 Budapest, Hungary}

\newcommand{\EXXON}{ExxonMobil Upstream Research Co., 3120 Buffalo Speedway, Houston, TX, 77098, USA}

\title{Avalanche dynamics on a rough inclined plane}

\author{Tam\'as B\"orzs\"onyi}
\email{btamas@szfki.hu}
   \affiliation{\LANL}
   \affiliation{\SZFKI}

\author{Thomas C. Halsey}
  \affiliation{\EXXON}

\author{Robert E. Ecke}
   \affiliation{\LANL}

\date{\today}

\begin{abstract}
Avalanche behavior of gravitationally-forced granular layers on a rough 
inclined plane are investigated experimentally for different materials and
for a variety of grain shapes ranging from spherical beads to highly 
anisotropic particles with dendritic shape.  We measure the front velocity,
area and the height of many avalanches and correlate the motion with the 
area and height. We also measure the avalanche profiles for several example
cases.  As the shape irregularity of the grains is increased, there is 
a dramatic qualitative change in avalanche properties. For rough 
non-spherical grains, avalanches are faster, bigger and overturning 
in the sense that individual particles have down-slope speeds $u_p$ that
exceed the front speed $u_f$ as compared with avalanches of spherical glass
beads that are quantitatively slower, smaller and where particles always
travel slower than the front speed.
There is a linear increase of three quantities 
i) dimensionless avalanche height ii) ratio of particle to front speed
and iii) the growth rate of avalanche speed with increasing avalanche size
with increasing $\tan\theta_r$ where $\theta_r$ is the bulk angle of repose, 
or with increasing $\beta_P$,  the slope of the depth averaged flow rule, 
where both $\theta_r$ and $\beta_P$ reflect the grain shape irregularity.
These relations provide a tool for predicting important dynamical 
properties of avalanches as a function of grain shape irregularity.
A relatively simple depth-averaged theoretical description captures some 
important elements of the avalanche motion, notably the existence of two 
regimes of this motion.
\end{abstract}


\maketitle

\section{Introduction}
\label{intro}

Granular materials form phases with strong similarities with ordinary 
phases of matter, i.e., solids, liquids or gases \cite{jana1996}. In 
many natural processes,  the coexistence of two of these phases is 
observed,  requiring a complex multi-phase description.
An example is avalanche formation, which occurs under various circumstances in
nature (snow avalanches, sand avalanches on dunes, rock avalanches,
land slides, etc.) as well as in industrial processes involving granular 
materials.  Although laboratory realizations of granular avalanches do 
not have the full complexity of avalanches encountered in nature, 
laboratory studies of this phenomena are important for understanding 
the fundamental behavior of avalanches where there is a interesting 
combination of stick-slip friction, yield criteria for the solid phase, 
and the fluid-like motion of the avalanche itself.  Although the
statistics of avalanche occurrences is a fascinating area of research 
\cite{fe1995}, we focus here on the dynamics of individual avalanche events.

There are several classes of experiments for avalanches, each with 
advantages and disadvantages.  One is to slowly rotate a closed cylinder 
that is about 50\% full of granular material 
\cite{ra1990,ra2002,boda2002,cogo2005}.  As the cylinder rotates, the 
angle of the granular surface exceeds the critical angle $\theta_c$, 
and material starts to flow along the surface.  At rapid rotation rates 
the flow is continuous but for slower rotation rates, avalanches occur 
because the rate of depletion of the granular material in the avalanche 
brings the surface back to the angle of repose $\theta_r$ faster than 
the rotation can maintain a surface angle greater than $\theta_c$.  
Significant advantages of this approach are that there is no need to 
continuously supply grains to the system, and the flow rate is easily
controlled by the rotation rate. Further, using transparent boundaries 
enables direct observation of the velocity profile in the flowing layer 
because avalanches in this system typically extend to the side walls. 
From such optical measurements one finds that the vertical velocity 
profile of an avalanche follows an exponential decay \cite{cogo2005} 
in contrast to steady flows where it has an upper linear part. On the 
other hand, the role of friction at the side boundaries on the 
vertical velocity profile is unclear and may be quite important because
the horizontal velocity profile is a plug flow with two 
exponential boundary layers at the walls.

A second realization of avalanches is to add granular material to a 
heap near its peak to induce granular motion \cite{ledu2000,du2000,neag2003}.
The heap can be three dimensional or can be confined between rigid barriers,
typically transparent glass or plastic plates, to form a unidirectional flow.
For high input mass flux near the top, the grains flow continuously down the 
surface formed by other grains. For lower incoming mass flow, avalanches form
intermittently.  The transition between continuous and avalanching flow as a 
function of input mass flux is hysteretic \cite{ra1990,liha1999}.
Studies using diffusing-wave spectroscopy \cite{du2000} and molecular
dynamics simulations \cite{si2005} provided detailed properties of this 
intermittency. The experimental results show that the microscopic grain dynamics 
are similar in the continuous and intermittent flow regimes \cite{ledu2000}.  
In the continuous flow regime for a bulk heap,  the grain velocity $u(z)$
decreases linearly as a function of the depth $z$ below the surface. 
Below some depth, however, a much slower ``creep" motion is observed with 
an exponential velocity profile \cite{koin2001,boda2002,ando2001}.
For the case of avalanches, theoretical results suggest that the 
avalanche amplitude should depend strongly on the velocity profile 
\cite{arra1999,emcl2005}.

\begin{figure}[ht]
\resizebox{70mm}{!}{
\includegraphics{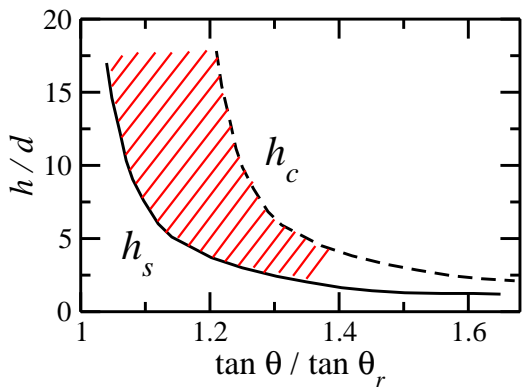}
}
\caption{(Color Online) The minimum layer thickness $h_s$ (averaged for 8 
different materials) at which flow stops (solid line) and an illustrative 
curve for the critical thickness $h_c$ at which flow starts (dashed line).
The operating conditions for avalanches (for the present work) fall
in the region represented by the hatched zone.
}
\label{hs-hc}
\end{figure}

A third laboratory system for avalanche studies is a layer of 
grains on a rough plane inclined at an angle $\theta$ with
respect to horizontal
\cite{dado1999,da2000,da2001,ivre2000,arts2001,ra2002aug,feth2004,mala2006,clma2007,arma2006}. 
Because we use this system in the experiments reported below, we 
review some aspects of these inclined layer flows in more detail. 
Grains on a rough inclined plane form a thin stable static layer 
even for $\theta >\theta_c$ and the onset of the flow is expected 
only above a critical thickness $h_c$, while the flow subsides at 
$h_s < h_c$ \cite{pona1996,po1999}. The values of $h_c$ and $h_s$ 
are rapidly decreasing by increasing $\theta$ as is schematically
illustrated in Fig. \ref{hs-hc}. Here $h_s$ is an average of the 
curves measured for 8 different materials (for the data see Section
\ref{results}), and $h_c$ is an approximate curve for illustration.

According to Bagnold \cite{ba1954},  the shear stress in granular 
flows is proportional to the square of the strain rate. This hypothesis 
has been checked in various configurations and has been proven to work
relatively well for dense inclined plane flows \cite{gdrmidi2004}. 
The Bagnold stress-stain relationship leads to a velocity profile
of the form $u(z) = u_0 (1-(h-z)/h)^{3/2}$ where the surface 
velocity $u_0$ depends on $h$ as $u_0\sim h^{3/2}$.
The above convex velocity profile is recovered in MD simulations for 
relatively thick flows \cite{sier2001,sila2003}. As the internal velocity
profile is difficult to determine experimentally, the most straightforward 
experimental test of the Bagnold hypothesis is to measure the height 
dependence of the surface velocity $u_0$ or the depth-averaged velocity 
$\overline u$. 
This relationship, called the flow rule (FR), was obtained for homogeneous 
flows with glass beads or sand \cite{po1999,fopo2003,boec2007} and is found 
to be consistent with the Bagnold flow profiles.  Two slightly different 
forms of the FR are the Pouliquen flow rule
\begin{equation}
\overline u/\sqrt{gh} = \beta_P h/h_s(\theta) - \gamma
\label{PFR}
\end{equation}
and a modified form due to Jenkins \cite{je2006} which we denote the 
Pouliquen-Jenkins flow rule:
\begin{equation}
\overline u/\sqrt{gh} = \beta_{PJ} (h/h_s(\theta))(\tan^2\theta/\tan^2\theta_1) 
\label{PJFR}
\end{equation} where $\theta_1$ is the vertical asymptote of the $h_s(\theta)$ curve.

In the inclined layer system, one can either prepare the
layer in a metastable state and mechanically induce a single avalanche 
\cite{dado1999,da2000,da2001,arts2001}, or an intermittent series of
avalanches can be produced by slowly adding grains near the top of 
the inclined layer as is done here. In the former case the whole layer 
becomes metastable \cite{dado1999,da2000,da2001}, and the shape and 
propagation of the avalanches depends critically on $\delta \theta$  
(the level of metastability), giving rise to downward and also upward 
expanding avalanches when triggered with a small perturbation.
For both kinds of avalanches on an inclined plane the angle at which flow 
starts or stops depends on the thickness of the layer contrary to the
heap experiments, on the properties of the bottom boundary such as 
surface roughness (which controls the zero-velocity boundary condition), 
and on the particular granular particle properties including inter-particle 
friction and particle shape. The distinct advantages of this system for 
the study of avalanches include the robust flow rule relationship in the 
continuous flow regime  \cite{po1999,anda2002,boec2007} that relates 
depth-averaged velocity and height, a simpler vertical velocity profile, 
and the ability to adjust the stability of the layer by changing the plane 
inclination angle $\theta$.  Note that both flow rules connect the rheology 
deep in the flowing state with properties of the boundary between the 
flowing and the stationary states.

The majority of laboratory experiments use spherical beads - an idealized
granular material.  However,  properties of waves, either on a rough incline
\cite{fopo2003} or in hopper flows \cite{babe1989}, depend on the shape
anisotropy of the grains.  Also, in recent numerical simulations the velocity 
profile in Couette flow depends strongly on the angularity of the particles
\cite{cl2008}. Thus, it is natural to expect that avalanches will
be similarly affected by the shape of the individual granular particles.  Indeed, 
glass beads and sand  show qualitatively {\it different} avalanching behavior
\cite{boha2005} despite qualitatively {\it similar} flow rules in the steady flow phase. 
Avalanches formed by sand particles are larger with more dynamic
grain motion than avalanches formed by glass beads. By plugging the known flow
properties \cite{fopo2003} into the depth-averaged model equations these
basic differences can be explained \cite{boha2005}.

In this paper, we present the results of an extensive study, using a set of different
materials, and describe the details of our experimental methods used for the
results presented in \cite{boha2005}. 
Our aim is to capture how the dynamical properties of the avalanches depend on 
the grain shape irregularity, including the effects due to deviations
from the idealized spherical shape of the overall grain form and effects
of the microscopic surface roughness of the particles.
We find relations by which we can quantitatively predict major properties
of avalanches as a function of the angle of repose $\theta_r$ or the slope of the flow rule
$\beta_P$, both of which provide a measure of grain shape irregularity.

We study the case where the layer is initially stable to small perturbations. 
As new grains are added to the layer at the top section ($5\%$) of the plane, 
the granular layer becomes locally unstable in a cyclic manner leading to the
intermittent formation of avalanches. These avalanches propagate down the plane 
on top of the stable static layer in a stationary manner, i.e., the shape and 
velocity of the avalanche does not change appreciably. The dynamical properties 
of these avalanches are studied far down the plane from where they are formed. 
We use eight different materials including rough irregular shaped particles of 
different sizes and spherical beads. The materials are characterized in 
Sec.\ \ref{exp} where the experimental conditions are also described. 
In Sec.\ \ref{results} we discuss how the properties of the avalanches depend 
dramatically on the shape (spherical, or irregular) of the grains. 
We describe a simple theory of avalanche flow for glass beads in Sec.\ \ref{theory}. 
Conclusions are drawn in Sec.\ \ref{conclusion}.

\section{Experimental section}
\label{exp}

In this section we describe the experimental apparatus and the 
characteristics of the granular materials.
The experimental techniques and apparatus have been described in 
detail elsewhere \cite{boha2005,boec2006} so only essential features 
are presented here.

\subsection{Experimental setup}

\begin{figure}[ht]
\resizebox{80mm}{!}{
\includegraphics{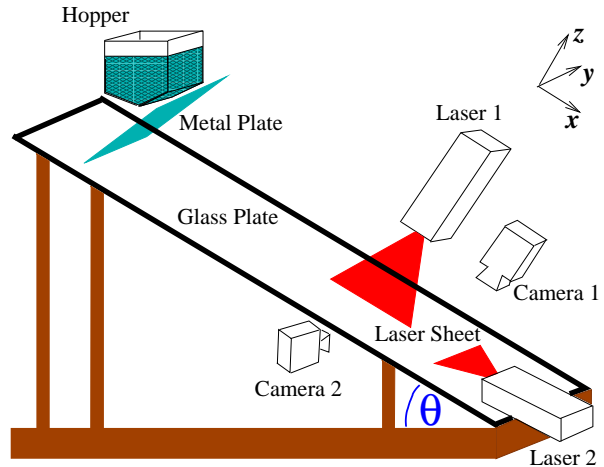}
}
\caption{(Color Online)
Schematic illustration of the experimental setup.
}
\label{setup}
\end{figure}

A sketch of the experimental setup is shown in Fig.\ \ref{setup}.
 The granular material flows out of the hopper at a constant flow rate $Q$.
The grains first hit a small metal plate that disperses the material
so that the mass flux per unit width $F=Q/W$ is relatively homogeneous in 
the {\it y} direction transverse (of total width $W$) to the inclination 
direction $x$.  The  grains are deposited on the top $5\%$ of the plane. 
The metal plate helps reduce possible electric charging of the particles 
during the hopper flow.  The granular layer reaches a critical state locally 
and an avalanche is formed and travels down the plane. The layer into which 
the avalanche moved has approximate thickness $h_s$.  The stable static 
layer is prepared by releasing grains at the top of the plane and letting 
the system relax, or by letting the hopper run for a longer period of time.
The glass plate has dimensions of 220 cm x 40 cm with a rough surface
prepared either by gluing one layer of the same particles onto it or by 
using sandpaper with different roughnesses.
The two main control parameters of the system are the plane inclination angle
$\theta$ and the nature of the granular material used.
The incoming flux and the roughness of the plane could be also varied. As we
show in the following, within the limits presented below, variations in these
latter parameters do not influence the results presented in this paper.
The dynamics of the avalanches are recorded with a fast video camera 
(up to 2000 frames per second) about 150 cm below the incoming flow (Camera 1).
The vertical profiles
of the avalanches along their symmetry axes are recorded with Camera 2
using a vertical laser sheet to detect the height differences (Laser 1).
With another laser sheet (Laser 2)  slightly inclined with respect to the 
glass plate, 2D height profiles of avalanches are reconstructed.

The incoming mass flow rate (per unit width along $y$) $F_0$ is fairly constant
for all measurements of one material and is in the range
$0.02 \leq F_0 \leq 0.2$ g/cm-s depending on the material. The value of $F_0$
is chosen by determining the flux $F_w$ needed to produce a wave state where
the entire width of the layer is in motion.
The properties of such waves were explored experimentally for glass beads
\cite{da2001}.  The flux $F_0$ is set to approximately $F_w/2$ to obtain a
reasonable number of distinct avalanches in the measurement area.  
For $F \leq F_w$, adjusting $F_0$ only affects the total number of avalanches
and their size distribution rather than the individual avalanche
characteristics that are the focus of this paper.

\subsection{Characterization of the materials}

\begin{figure}[ht]
\resizebox{83mm}{!}{
\includegraphics{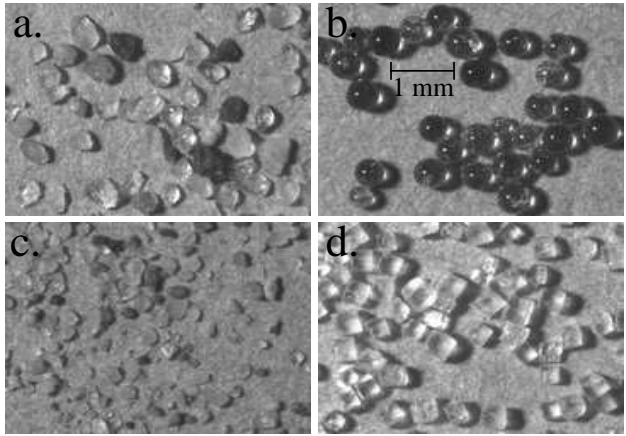}
}
\caption{ Microscopic images of the materials: (a) sand with diameter
$d=400\pm100\ \mu m$, (b) spherical glass beads with $d=500\pm100\ \mu m$,
(c) fine sand with $d=200\pm100\ \mu m$ and (d) salt with $d=400\pm100\ \mu m$.
}
\label{samplesgran}
\end{figure}

The shape of the granular particles has a large impact on the
qualitative and quantitative behavior of avalanches in our system whereas
other properties such as the mean particle diameter $d$ have less effect.
For example, despite some relatively low polydispersity for the materials
investigated ($\leq \pm 30\%$ except for fine sand which had $\pm 50\%$),
we do not observe any effects attributable to size-induced segregation.
To show the nature of the shape anisotropy qualitatively, we first show
in Fig.\ \ref{samplesgran} images of the glass beads ($d=400\pm100\ \mu m$),
fine ($d=200\pm100\ \mu m$)  and coarse ($d=400\pm100\ \mu m$) sand and salt
particles ($d=400\pm100\ \mu m$).
The glass beads are quite spherical although no attempt has been made 
to eliminate slightly aspherical particles, and no quantitative analysis 
of asphericity has been done.  The sand has many irregular shapes ranging 
from approximately spherical to very angular. Finally, the salt grains 
take on very angular shapes reflecting the cubic symmetry of salt crystals.
For the purposes of our studies, we assume that the grains do not change 
appreciably with time because the number of realizations of flows is limited,
and the particle velocities are small. Thus, grains are not subjected to 
the repetitive collisions that might smooth their shape thereby changing 
the shape anisotropy or inter-particle friction.
\begin{figure}[ht]
\resizebox{83mm}{!}{
\includegraphics{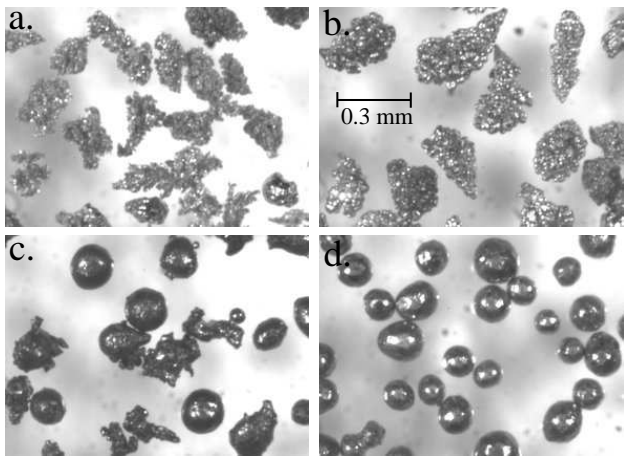}
}
\caption{ Microscopic images of the copper particles of size
$d=160\pm50\ \mu m$ and with packing fractions
(a)  $\eta=0.25$,
(b)  $\eta=0.33$,
(c)  $\eta=0.5$
and (d)  $\eta=0.63$.
}
\label{copperimages}
\end{figure}
The copper materials are used to explicitly introduce a controlled shape 
distribution and explore the consequences of very different shapes on 
avalanche behavior.  In Fig.\ \ref{copperimages}, we show images of commercial
(ECKA Granules GmbH $\&$ Co.,KG.) copper particles (mean diameter 
$d \approx 160\pm50\ \mu$m) with very different average shapes.
The particles range from highly irregular, dendritic-like shapes to almost 
spherical.  The overall anisotropy is measured by the static packing fraction 
$\eta$ for each shape, ranging from 0.25 for irregular particles to 0.63 for 
spherical copper particles.  For comparison, the packing fraction for sand is 
about 0.56 and for spherical glass beads is 0.63.

\begin{figure}[ht]
\resizebox{85mm}{!}{
\includegraphics{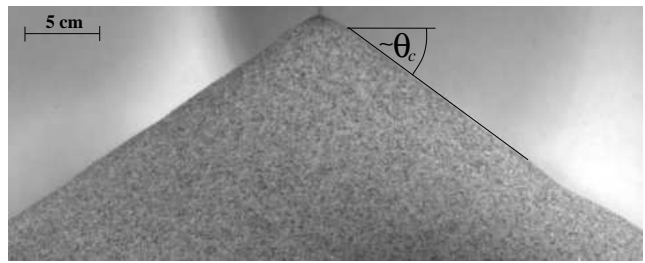}
}
\caption{Image of a sandpile - a heap of coarse sand particles.  
The scale is indicated in the figure and the solid line indicates the local 
slope used to determine $\theta_c$ as shown or more generally $\theta_r$ as well.
}
\label{sandpile}
\end{figure}

The next thing to consider is how the different shapes and properties of the
materials affect the basic granular flow properties, in particular
$\theta_r$ and $\theta_c$ of the bulk material.  We do so by 
observing avalanches on bulk heaps with a high speed video camera. 
As material is added at a very small rate
to the top of the pile, avalanches form and propagate downwards
intermittently. The basic configuration is shown in Fig.\ 
\ref{sandpile} where a bulk pile of sand is shown with a line 
indicating the determination of local $\theta_c$ (just before an 
avalanche) or $\theta_r$ (just after an avalanche, not shown).

\begin{figure}[ht]
\resizebox{85mm}{!}{
\includegraphics{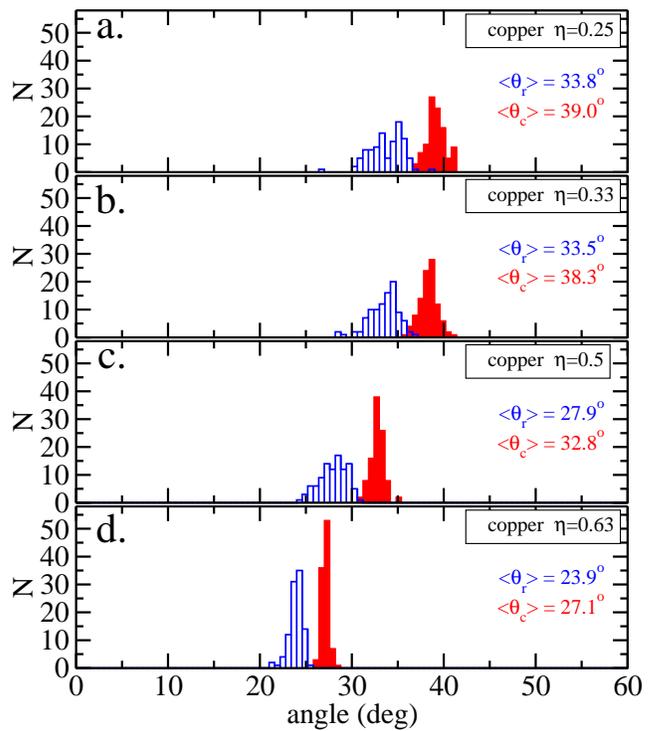}
}
\caption{(Color Online) Distributions of the critical angle 
$\theta_c$ (filled columns)
and the angle of repose $\theta_r$ (open columns)
measured for 100 avalanches on a three dimensional sandpile for
copper with packing fractions
(a)  $\eta=0.25$,
(b)  $\eta=0.33$,
(c)  $\eta=0.5$
and (d)  $\eta=0.63$. Particle sizes $d=160\pm50\ \mu m$.
}
\label{reposehistogramcopper}
\end{figure}

The distribution of the critical angle $\theta_c$ and that of the 
angle of repose $\theta_r$ is plotted for 100
avalanches for each of the different materials investigated.  
\begin{figure}[ht]
\resizebox{85mm}{!}{
\includegraphics{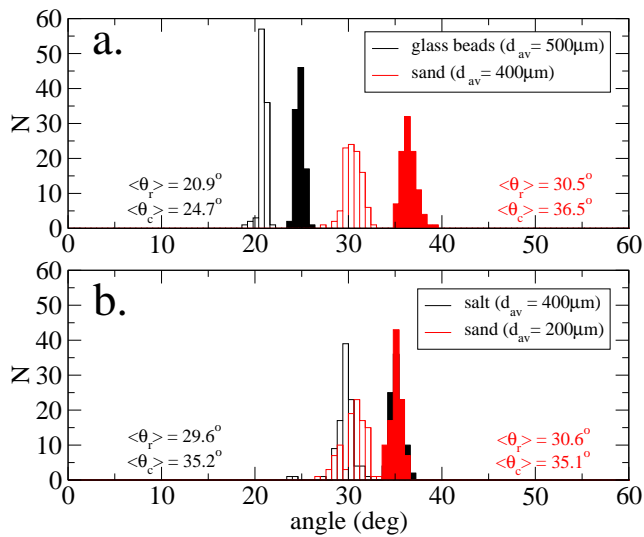}
}
\caption{ (Color Online) Distributions of the critical angle $\theta_c$ 
(filled columns) and the angle of repose $\theta_r$ (open columns) measured 
for 100 avalanches on a three dimensional sandpile for (a) sand (red) and 
glass beads (black), (b) fine sand (red) and salt (black).}
\label{reposehistogram}
\end{figure}
The distributions for the glass, sand and salt are shown in
Fig.\ \ref{reposehistogram} and the distributions for the copper 
particles are presented in Fig.\ \ref{reposehistogramcopper}.
The average values of $\theta_c$ and $\theta_r$ are indicated.
As expected, $\theta_c$ and $\theta_r$ are significantly
higher for the piles formed by particles of irregular shape compared 
to piles of spherical particles. The widths of the distributions are 
also larger for irregular particles because shape irregularity gives 
rise to a larger variety of configurations and a wider range of angles 
at which avalanches start or stop.

For thin layers on an inclined plane, grains start or stop flowing at angles
determined by the layer thickness $h$ as well as by their individual properties
such as shape or surface roughness.  In our experiments, we determine for each
material the height $h_s(\theta)$ at which the material stops flowing, the thin
layer equivalent of the bulk angle of repose.  In a technique described in detail
elsewhere \cite{boec2007}, we allow material to flow by slowly adding grains at
the top of the plane resulting in intermittent avalanches.
Upon stopping the input, the layer comes to rest, and the volume of grains on the
whole plane is determined. Knowing the surface area of the plate allows a
determination of the mean height $h_s$ with high precision, typically 2$\%$.
Alternatively, a large quantity of grains is placed on the inclined plane, and a
continuous flow persists over the entire plane for 10 - 20 seconds. After the flow
subsides, the thickness of the resulting static layer is the same to within
experimental error as that determined by the first method.

\begin{figure}[ht]
\resizebox{85mm}{!}{
\includegraphics{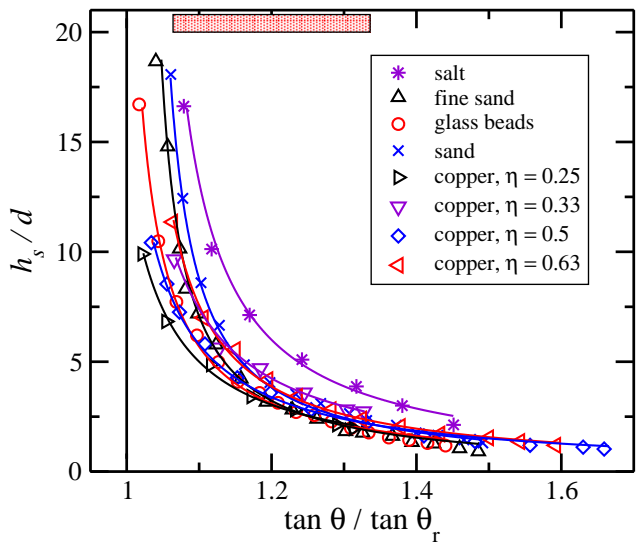}
}
\caption{(Color Online) Static layer thickness normalized by $d$ as a function
of $\tan\theta$ normalized by $\tan\theta_r$. The continuous lines are best 
fits to the formula $h_s = a d / (\tan \theta - \tan \theta_1)$
with the resulting fitting parameters $\alpha$ and $\theta_1$ indicated in
Table \protect \ref{parameters}. A vertical solid line indicates 
$\theta_r$. The horizontal bracket near the top of the figure indicates 
the range of $\theta$ over which avalanches are measured.
}
\label{staticlayer}
\end{figure}

Many of the details of the characteristics of flow on our inclined plane, 
including the detailed comparison with the Pouliquen flow rule \cite{po1999} 
and the effects of surface roughness, have been presented elsewhere \cite{boec2007}. 
We summarize our results for the particular materials used here by collapsing 
the data using a normalized ratio of $\tan\theta/\tan\theta_r$ where $\theta_r$
is determined from the bulk measurement and normalizing the height by the particle 
diameter $d$, see Fig.\ \ref{staticlayer}.  
The continuous lines are best fits to the formula
$h_s = a d / (\tan \theta - \tan \theta_1)$.
The resulting values for the fitting parameters $\alpha$ and $\theta_1$ are
indicated in Table \ref{parameters}.
For all the materials, the stopping height $h_s$ decreases with increasing
$\theta$ as shown in Fig.\ \ref{staticlayer}, dropping rapidly for angles
close to the bulk angle of repose and more gradually for larger $\theta$.
\begin{table}[h]
\caption{\label{parameters} Fitting parameters $a$ and 
$\theta_1$, resulting as
best fits to the data in Fig. \ref{staticlayer} using the formula
$h_s = a d / (\tan \theta - \tan \theta_1)$.}
\begin{tabular}{|c|c|c|c|}
\hline
      sample & $a$ & $\theta_1$ \\
      \hline
           \hline
      salt & 0.63 & $30.0^{\circ}$ \\
      fine sand & 0.35 & $31.0^{\circ}$ \\
      glass beads & 0.26 & $20.5^{\circ}$ \\
      sand & 0.4 & $31.1^{\circ}$ \\
      copper, $\eta=0.25$ & 0.66 & $32.6^{\circ}$ \\
      copper, $\eta=0.33$ & 0.52 & $32.3^{\circ}$ \\
      copper, $\eta=0.5$ &  0.43 & $26.9^{\circ}$ \\
      copper, $\eta=0.63$ & 0.35 & $23.8^{\circ}$ \\
      \hline
\end{tabular}
\end{table}
The curves in Fig.\ \ref{staticlayer} are for sandpaper with a roughness 
of $R=0.19$ mm. The corresponding data curve for sand on a surface
prepared by glued sand particles is identical within experimental error.
The collapse of the data suggest that differences in plane
roughness do not translate into significant changes in $h_s$.
We have also tested the influence of the incoming flux and kinetic energy
of the incoming particles on $h_s$.  For all the above configurations tested
using different initial flow rates, the data points fall on the same curve
within $\pm 5 \%$ for $\tan\theta/\tan\theta_r>1.1$. When approaching 
$\theta_r$ the measurements become less accurate with the rapid increase 
of $h_s$ leading to a $\pm 12 \%$ uncertainty of the data points for
$\tan\theta/\tan\theta_r<1.1$.

There are several additional remarks about the system that are important.
First, the majority of our measurements do not require detailed information 
about $h_c(\theta)$, the thickness at which grains start to flow for a 
particular $\theta$.  Thus, we have not made measurements of $h_c$. When 
data about this quantity are needed as in the modeling section presented below, 
$h_c(\theta)$ is estimated from measurements in similar systems \cite{po1999}. 
Second, avalanches are observed in the range of 
$1.04 <\tan \theta/\tan \theta_r< 1.34$ as indicated by the bar on
the top of Fig. \ref{staticlayer}. For higher $\theta$, even if a homogeneous
static layer is prepared beforehand, the kinetic energy of the incoming 
grains is enough to slowly erode the pre-existing layer, and avalanches can 
only be observed for a short time.

\subsection{Experimental procedures and analysis}

\begin{figure}[ht]
\resizebox{85mm}{!}{
\includegraphics{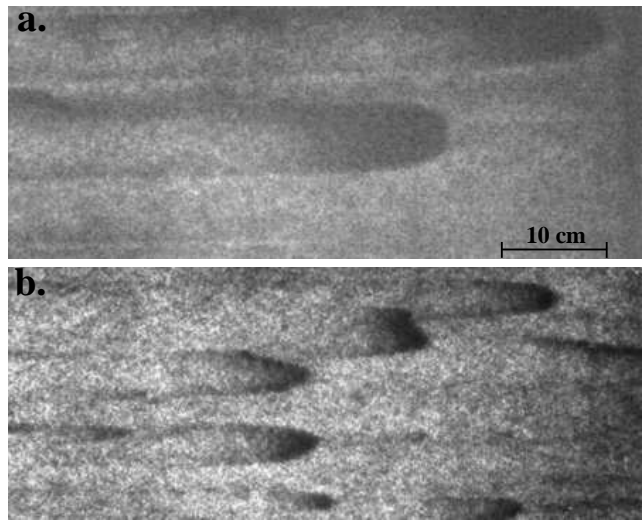}
}
\caption{Images (a) at $\theta=33.6^\circ$ and (b) at $\theta=38.1^\circ$
using transmitted light taken with Camera 1 for sand with $d=400\mu$ m.
}
\label{avalimages}
\end{figure}

For different materials and for different angles $\theta$, we produce 
avalanches by adding grains continuously at the top of the plane. The 
main quantities of interest are the velocity of the front or of the individual 
grains, the lateral extent of the avalanche measured by the avalanche area 
$A$, and the height of the layer $h$ as a function of position.  These quantities 
are measured by analyzing images taken with the high-speed video cameras.  
Two images taken of transmitted light (Camera 1) for coarse sand at 
$\theta=33.6^\circ$ and $\theta=38.1^\circ$ are shown in Fig.\ \ref{avalimages}.
These avalanches are localized objects that can be isolated
by differencing of subsequent images because regions were the
grains move become visible on a uniform background. The area of the
avalanche is determined by the areal extent over which finite displacements
in image differencing are measured.  Similarly the location of the front of
the image differencing region is used to determine the front velocity $u_f$.

\begin{figure}[ht]
\resizebox{85mm}{!}{
\includegraphics{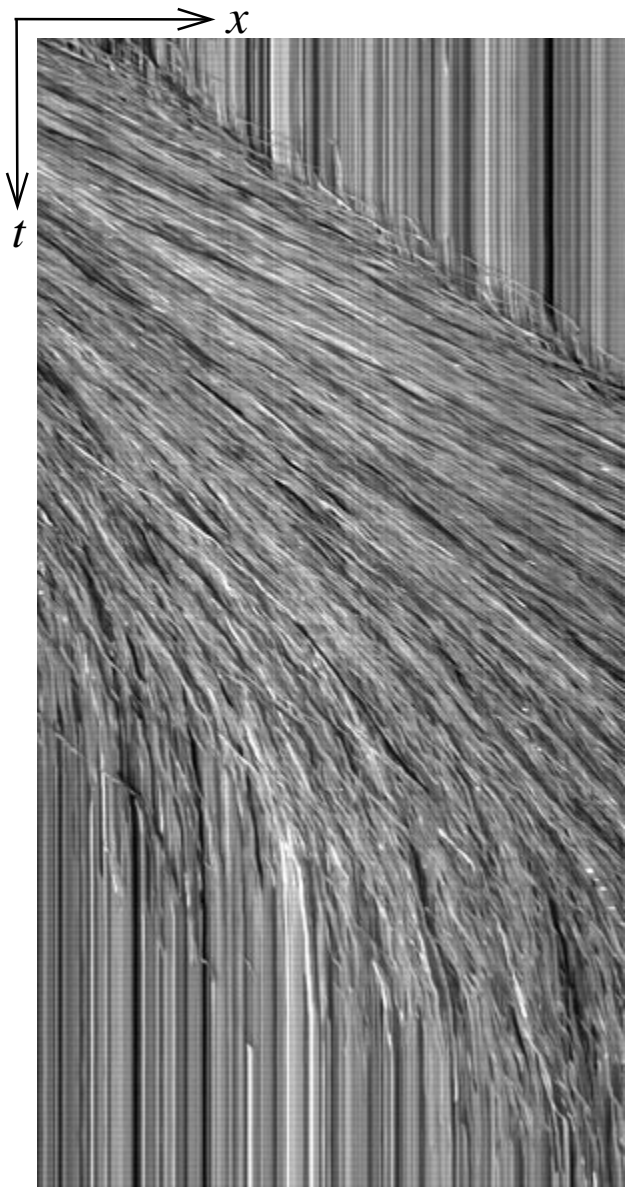}
}
\caption{Space-time plot for a sand avalanche for $\theta=36.8^\circ$ 
and $h_s=0.12$ cm.
Image size: 5.96 cm x 1.43 s. (Velocity profile shown in Fig.\ \ref{height-veloprofiles}(a).)
}
\label{spatiofull}
\end{figure}

To track the motion of individual grains inside the avalanches, a sequence of
images are extracted from movies of the granular flow dynamics \cite{movieweb}.
Space-time plots are created by taking the intensity along the symmetry axis of
the avalanche from movies taken by Camera $1$.  An example is shown in
Fig.\ \ref{spatiofull} for sand.  
The vertical lines starting on the top of the diagrams are the traces of 
particles on the top of the static layer in front of the avalanche.  
The tilted trajectory dividing this region from the dynamic region is the 
trace of the avalanche front, providing a second method for determining $u_f$. 
The traces of moving individual grains inside the avalanche are tilted lines.  
The change in the tilt of these traces as we go downwards (towards the end of 
the avalanche) indicates the decreasing particle velocity $u_p$ behind the front.
By analyzing the space-time plots the mean particle velocity $u_p$ has been 
measured at several locations and the velocity profile $u_p(x)$ has been 
determined. For this example using sand particles, the particles in the 
avalanche core (close behind the front) move faster than the front velocity 
of the avalanche.

\begin{figure}[ht]
\resizebox{80mm}{!}{
\includegraphics{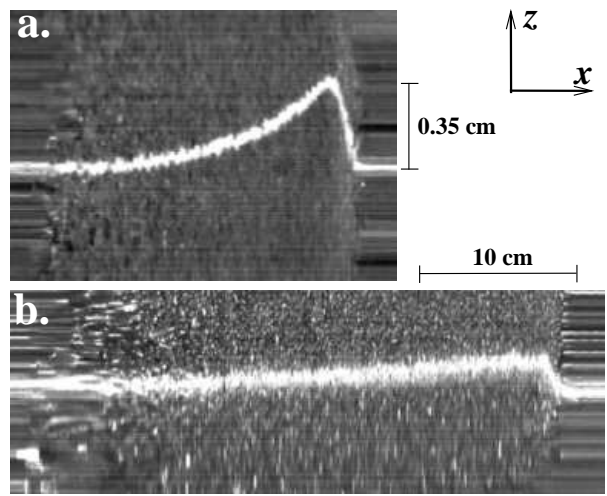}
}
\caption{Avalanche profiles taken from the side with the help of a laser
sheet for (a) sand (for $\theta=33.6^\circ$) and (b) glass beads
(for $\theta=22.6^\circ$). The horizontal size is contracted by a factor of 25.
}
\label{laserprofiles}
\end{figure}

In order to trace the height profile of avalanches, a vertical laser sheet
($xz$ plane) is projected onto the plane. Movies are taken of avalanches
that are cut by the laser sheet near their center (symmetry axis).
The camera is mounted for these experiments on the side at an angle of about
$13^\circ$ with respect to the $xy$ plane (Camera 2 in Fig.\ \ref{setup}).
By measuring the speed of the avalanche and taking the intensity only along 
one vertical line of these movies, the profile of the avalanches can be 
traced assuming that the profile does not change as it passes through.  
In Figs.\ \ref{laserprofiles}(a) and \ref{laserprofiles}(b), profiles 
are shown for a sand and for a glass avalanche, respectively. The images 
are contracted by 25 times in the horizontal direction. This technique 
allows the instantaneous profile along one line to be determined.

\begin{figure}[ht]
\resizebox{80mm}{!}{
\includegraphics{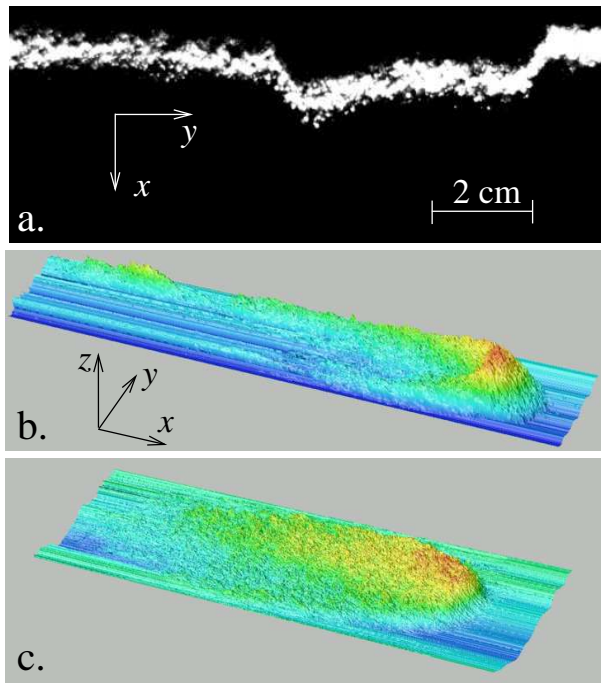}
}
\caption{
(Color Online) a) Image of the laser profile taken with Camera 1.
The height of the avalanche is obtained as
$h = h_s+\delta x \tan  \phi$, where
$\phi$ is the angle between the plane and the laser sheet (Laser 2)
and $\delta x$ is the displacement of the laser line.
Height profiles of (b) sand avalanche for $\theta=36.8^\circ$.
Image size $6.9$   cm  x  $34.6$ cm, (vertical size rescaled by 25x)
maximum height: $h_m=0.35$  cm, static layer thickness $h_s=0.12$  cm;
(c) glass bead avalanche for $\theta=24.3^\circ$.
Image size $10.3$  cm  x  $38.4$  cm, (vertical size rescaled by 25x) maximum
height: $h_m=0.28$   cm, static layer thickness $h_s=0.18$  cm.
}
\label{avalanche3D}
\end{figure}

Another way to visualize the avalanches is to trace a laser line (Laser 2),
see Fig.\ \ref{setup}, across the direction of avalanche propagation, i.e.,
transverse to the plane inclination direction.  Using the deflections of 
this laser line,  the whole 2D surface of the avalanche is obtained. A sample 
image (from Camera 2) is shown in Fig.\ \ref{avalanche3D}(a) for a sand avalanche.
 From the image sequence, the 2D height profile is reconstructed and is
shown in Fig.\ \ref{avalanche3D}(b) for the same avalanche and similarly
for an avalanche formed by glass beads in Fig.\ \ref{avalanche3D}(c).

\section{Results}
\label{results}

The main objective of our experimental investigation is to understand 
the similarities and differences between avalanches of granular materials 
with different properties, particularly grain shape irregularity.  
We first describe some of the qualitative characteristics of the avalanches
and then provide quantitative measures of those properties.

The avalanches we investigated are stationary (do not change with time) 
above a certain size with an overall shape that depends on the granular material.
Glass bead avalanches have an oval shape, Fig.\ \ref{avalanche3D}(c), whereas
medium sized or large sand avalanches have two tails, Figs.\ \ref{avalimages}(b)
and \ref{avalanche3D}(b), which can break off and form very small avalanches. 
Small sand avalanches typically do not have these tails. Very small avalanches 
are not stationary, but decelerate, lose grains, and eventually come to rest.
The length of the observation area is about 8-12 times the length of small
avalanches. Avalanches that are decelerating or that stop
in the field of observation are not included in the data presented here.
Interacting/merging avalanches are also eliminated.

\begin{figure}[ht]
\resizebox{85mm}{!}{
\includegraphics{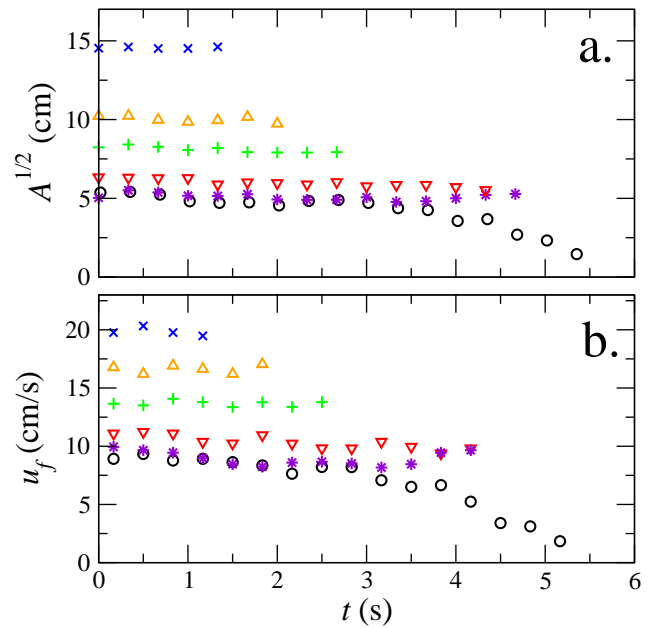}
}
\caption{(Color Online)
Time evolution of the (a) lateral size $A$ and (b) the front
velocity $u_f$ of four avalanches for sand with $d=400\ \mu$m and
$\theta=35.2^\circ$.
}
\label{size-time-velo-time}
\end{figure}

To demonstrate the stationarity of avalanche size and speed, the time 
evolution of the front velocity $u_f$ and the characteristic size 
(the square root of the lateral area $A$) is shown for a set of avalanches
in Figs.\ \ref{size-time-velo-time}(a) and \ref{size-time-velo-time}(b)
for sand taken at $\theta=35.2^\circ$. The smallest, slowest moving 
avalanches have a discernibly downward slope indicating a shrinking, 
decelerating avalanche.  If the decay is exponential, the linear slope 
would yield a decay time of about 40 s for both the size and speed.  
For two avalanches of this initial size, one drops below a threshold, 
about 7 cm/s, and then quickly shrinks and decelerates until it vanishes.
Another with about the same initial conditions, decreases in size and 
speed for awhile but seems to recover and survive.

To understand the difference in avalanche survival for small sand avalanches, 
one needs to take account of the response of the static layer to the passage 
of an avalanche. Medium or big sand avalanches typically take material with 
them along their center and deposit grains along their edges (see Figs. 
\ref{avalimages} and \ref{avalanche3D}).  The change in the layer height is 
about $\pm$ 10$\%$ of $h_s$.  Thus, the different fates of austensibly similar
sand avalanches suggests that small differences in residual layer thickness 
influence their evolution in that a small avalanche that encounters a slightly
thinner region behind a recent large avalanche does not survive whereas a small
avalanche propagating along the ridge formed by a large avalanche picks up a bit 
of mass and manages to survive (at least over the size of the interrogation window).
No formation of residual ridges left by the wake of the avalanche is observed for
small sand avalanches or for avalanches formed by glass beads. 
Although we have not studied this effect in detail for all the materials, one 
might surmise that the residual wake structure is a property of irregular grain
avalanches whereas the spherical grain avalanches do not have this property.

\begin{figure}[ht]
\resizebox{80mm}{!}{
\includegraphics{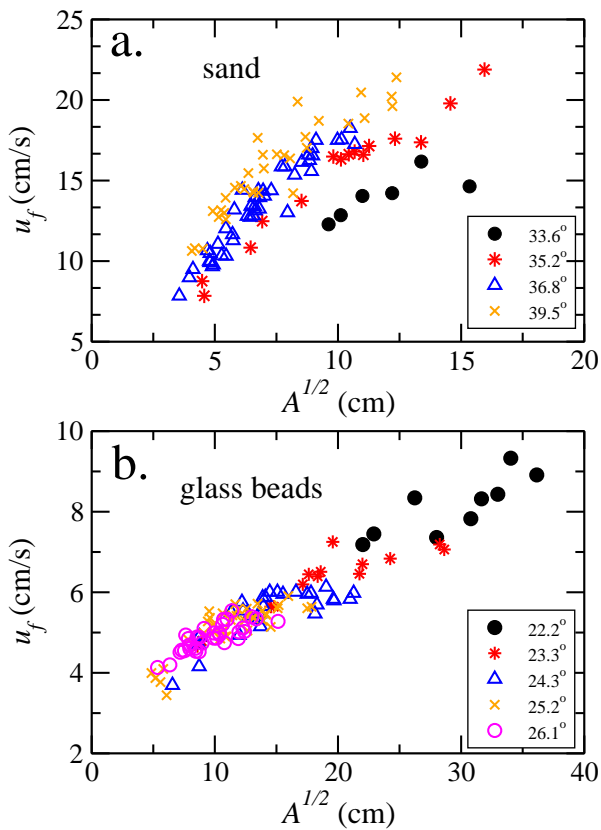}
}
\caption{(Color Online) Avalanche front velocity  $u_f$ as a function of the square root
of the lateral avalanche area $A$ for a)  sand and b) glass beads.
}
\label{size-veloSI}
\end{figure}

We now present a quantitative analysis of the speed and size of avalanches 
whose properties are stationary over the length of the channel, about 200 cm.
The avalanche velocity $u_f$ increases with avalanche size as shown in
Figs.\ \ref{size-veloSI}(a) and \ref{size-veloSI}(b) for fine sand 
and glass beads, respectively.
Similar sets of data were obtained for all eight materials. For anisotropic grains,
avalanches move faster with increasing $\theta$, see Fig. \ref{size-veloSI}(a).
On the other hand, avalanches formed by spherical beads are independent of $\theta$
in the sense that the curves obtained at various plane inclinations collapse, 
see Fig.\ \ref{size-veloSI}(b). This observation is not in contradiction
with data presented in \cite{da2001}, where avalanche speed for glass 
beads increases with decreasing $\theta$.
In fact, the tendency is the same here for the case of glass beads, as the average 
avalanche size increases with decreasing $\theta$, Fig.  \ref{size-veloSI}(b), 
so that bigger avalanches (with larger velocity) are observed for less steep inclines.
A similar analysis for the anisotropic particles is less conclusive,
owing to the wide range of sizes and velocities measured. 
Nevertheless, after averaging $u_f$ for all avalanches of different sizes, the
$u_f(\theta)$ curve for anisotropic grains is fairly independent of $\theta$ as well.

\begin{figure}[ht]
\resizebox{85mm}{!}{
\includegraphics{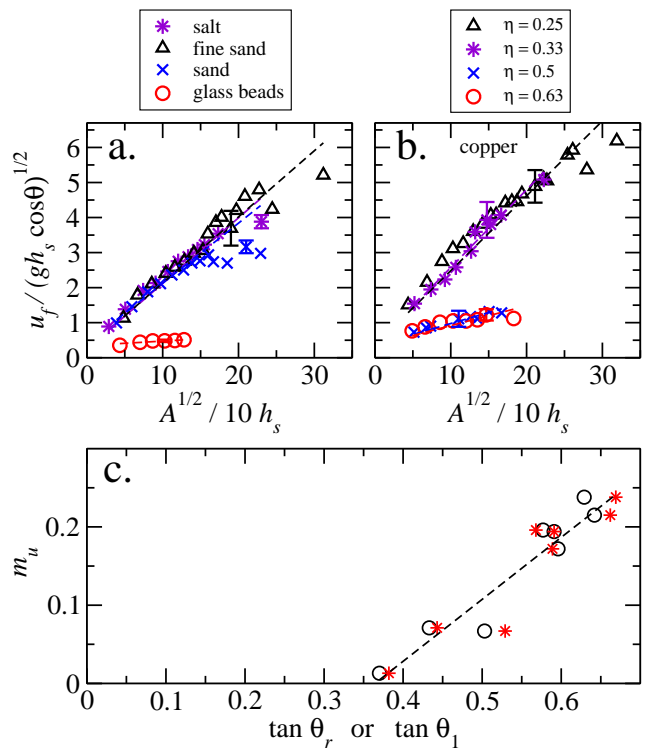}
}
\caption{(Color Online) Dimensionless avalanche velocity as a function of avalanche 
size for (a) or sand (x), glass beads (o), salt (*), fine sand ($\triangle$),
and (b) copper particles with $\eta = 0.25$ ($\triangle$),
$\eta = 0.33$ (*), $\eta = 0.5$ (x) and $\eta = 0.63$ (o).
The dashed lines are best fits to the data.
(c) Slope $m_u$ of the linear fits from (a) and (b) for all materials as a 
function of $\tan\theta_r$ (*) or $\tan\theta_1$ ($\circ$). 
}
\label{size-velo}
\end{figure}

To compare the $u_f(A^{1/2})$ curves for the different materials,  it is useful
to employ dimensionless parameters. For gravity driven flows on an
incline the relevant length scale is $h_s$ and the appropriate 
velocity scale is $\sqrt{g h_s \cos \theta}$ \cite{erha2002}.
The non-dimensionalization of the $u_f(A^{1/2})$ curves by these quantities
collapses the data taken at various plane inclinations for each material.
The collapsed curves are shown in Fig.\ \ref{size-velo}(a) for sand, salt,
and glass beads and in Fig.\ \ref{size-velo}(b) for the copper particles.
The dimensionless avalanche velocity increases linearly with increasing 
dimensionless avalanche size for all materials.  The data fall into two 
classes: irregular shaped particles form avalanches that
reach sizes about two to four times larger than those formed by spherical
beads. Further, the speed of the avalanches from the irregular grains is about
4 times larger than that of the avalanches formed by spherical beads. 
For a better characterization of these differences, we compare the slope 
$m_u$ of the best fits to the data (shown with dashed lines in 
Fig.\ \ref{size-velo}(a) and Fig.\ \ref{size-velo}(b)) for all materials.
In Fig.\ \ref{size-velo}(c) $m_u$ is plotted as a function of the tangent of the 
angle of repose $\theta_r$ (or $\theta_1$ the vertical asymptote of the
$h_s(\theta)$ curve), both of which provide a measure of grain shape irregularity.
We find that $m_u$ systematically increases with grain shape irregularity. 
In other words, the avalanche velocity increases with increasing avalanche 
size systematically for more irregular grains. Note that $m_u$ goes to zero 
at about $\tan\theta_r=0.36$, implying that for materials with a small angle
of repose (below about $\theta_r=19.8^\circ$) the dimensionless avalanche 
velocity is small (about $u_f/\sqrt{gh_s\cos\theta} \approx 0.35$)
and independent of avalanche size.

Another measure of the avalanche size is its height, in particular the maximum 
height $h_m$ along the avalanche profile.  We determine the height profile 
using the laser line method described in Sec.\ \ref{exp}.
In Figs.\ \ref{laserprofiles}(a) and \ref{laserprofiles}(b), we see 
qualitatively that the height of glass avalanches is considerably smaller 
than that of sand avalanches.  A second difference is that after the rapid 
increase in height at the front, a fast (exponential like) decrease is 
observed for sand avalanches whereas the height variation is straighter 
for avalanches formed by glass beads.

\begin{figure}[ht]
\vspace*{0.1cm}
\resizebox{85mm}{!}{
\includegraphics{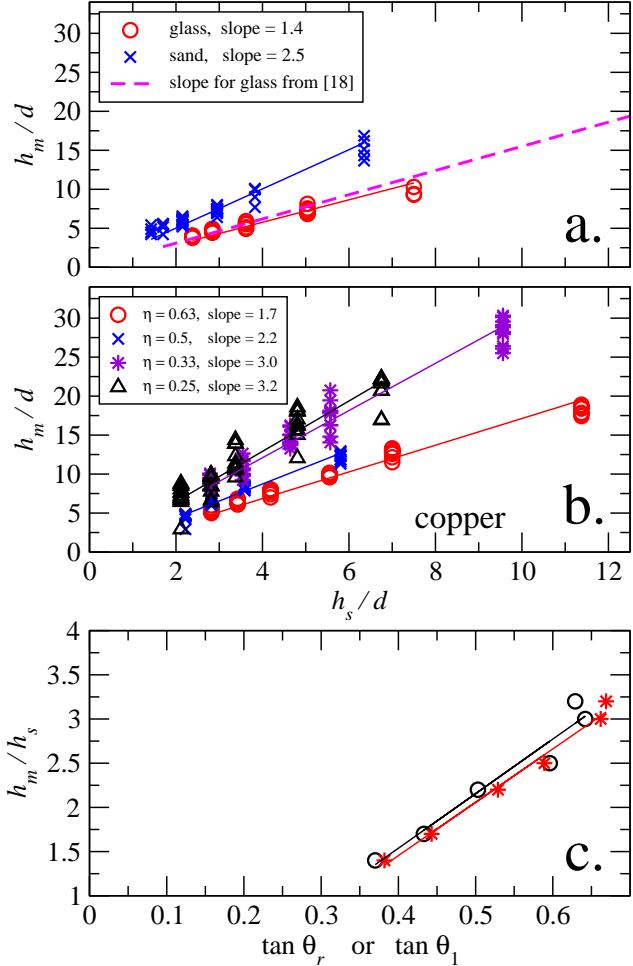}
}
\caption{(Color Online) Avalanche peak height $h_m$ as a function of the
static layer thickness $h_s$ for (a) sand (x) and glass beads ($\circ$), (b)
copper with $\eta = 0.25$ ($\triangle$),
$\eta = 0.33$ (*), $\eta = 0.5$ (x) and $\eta = 0.63$ (o).
(c) The ratio $h_m/h_s$ as a function of $\tan\theta_r$ (*) 
or $\tan\theta_1$ ($\circ$). The continuous lines are linear fits.  
}
\label{height-layerthickness}
\end{figure}
Numerous avalanche profiles have been recorded for sand, glass beads
and all copper samples at the same sequence of plane inclinations 
as before to provide a quantitative analysis of avalanche heights.
Plotting the avalanche peak height $h_m/d$ as a function
of the thickness of the underlying layer $h_s/d$, we find a 
systematic linear increase as shown for glass beads and sand in 
Fig.\ \ref{height-layerthickness}(a), implying that at lower plane 
inclinations where the static layer is thicker the avalanche height 
is proportionally larger. 
The slope $m_h$ of the curves gives a measure of the dimensionless
avalanche height $h_m/h_s$ for a given material which is basically
independent of the avalanche size and plane inclination and is
$m_h=1.45$ and $m_h=2.5$ for glass beads and sand, respectively. 
Also shown is the slope $m=1.55$ for glass bead waves on a velvet 
cloth \cite{da2001}. The close correspondence despite the considerable 
differences in systems (cloth versus hard, rough surface) and phenomena 
(waves versus avalanches) is striking.
For the copper particles (see Fig.\ \ref{height-layerthickness}(b)) $m_h$ falls 
in the range of $1.7 < m_h < 3.2$ as $\eta$ changes between $0.25 < \eta < 0.63$.
The general trend is that irregular particles form higher avalanches compared 
to avalanches of more spherical particles.
To quantify this trend we again plot $h_m/h_s$ as a function of the tangent 
of the angle of repose $\theta_r$ (or $\theta_1$ the asymptote of the
$h_s(\theta)$ curve). 
As seen in Fig.\ \ref{height-layerthickness}(c), $h_m/h_s$ increases 
systematically with $\tan\theta_r$. This is an important relation as it provides 
a tool for predicting typical avalanche heights (for a given material) just by 
measuring the angle of repose of the material.

Having demonstrated that both the dimensionless avalanche height $h_m/h_s$ 
and the growth rate of avalanche speed with increasing avalanche size 
(defined as $m_u$) depend systematically on increasing grain shape 
irregularity, we can further explore whether such a tendency
can be observed in other properties of avalanches.  Using the space-time 
technique described in Sec.\ \ref{exp}, we can explore the ratio of the 
mean particle speed near the front $u_p$ to the front velocity $u_f$.

\begin{figure}[ht]
\resizebox{85mm}{!}{
\includegraphics{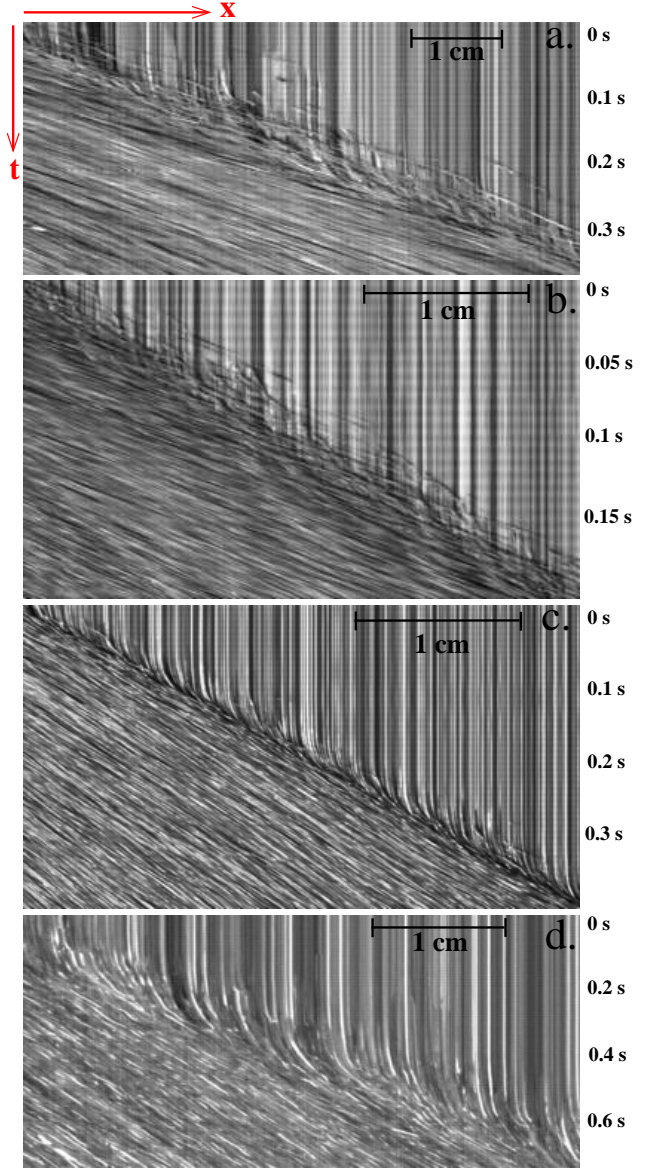}
}
\caption{Space-time plots taken at the symmetry axis of avalanches
with camera 1. (a) sand at $\theta=36.8^\circ$ ($\theta/\theta_c=1.01$),
(b) copper with $\eta=0.25$ at $\theta=39.5^\circ$ ($\theta/\theta_c=1.01$),
(c) copper with $\eta=0.63$ at $\theta=26.1^\circ$ ($\theta/\theta_c=0.96$)
and (d) glass at $\theta=23.5^\circ$ ($\theta/\theta_c=0.95$).
}
\label{spatio}
\end{figure}

\begin{figure}[ht]
\vspace*{0.1cm}
\resizebox{85mm}{!}{
\includegraphics{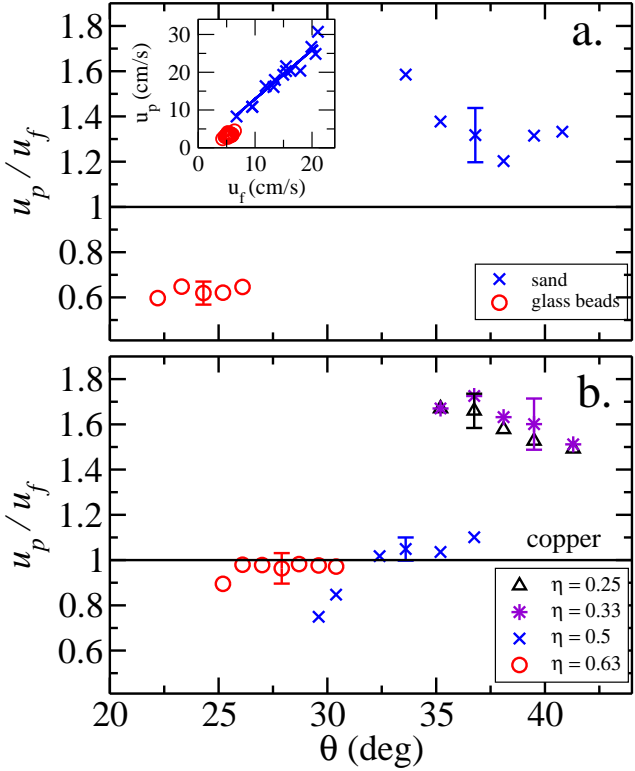}
}
\caption{(Color Online) Ratio of particle velocity $u_p$ and front velocity $u_f$
as a function of the plane inclination for (a) sand (x) and glass beads
($\circ$), (b) copper particles with $\eta = 0.25$ ($\triangle$),
$\eta = 0.33$ (*), $\eta = 0.5$ (x) and $\eta = 0.63$ (o).
The inset of (a) shows $u_p$ as a function of $u_f$ for sand (x)
at $\theta=36.8^\circ$ and glass beads (o) at $\theta=25.2^\circ$.
}
\label{velopart-per-front}
\end{figure}

To compare the behavior at the front, four examples are shown in
Figs.\ \ref{spatio}(a)-\ref{spatio}(d) for sand, copper with $\eta=0.25$,
  copper with $\eta=0.63$, and glass beads, respectively.
In Figs.\ \ref{spatio}(a) and \ref{spatio}(b) one sees that for anisotropic
particles the velocity of particles is larger than the front velocity;
in many cases particles are flying out of the main body of the avalanche
and are stopped by the static layer in front of the avalanche.
On the contrary for the avalanches formed by spherical beads,
Figs.\ \ref{spatio}(c) and \ref{spatio}(d) particles are slower than 
the avalanche front, and particles in the static layer begin moving just 
before the actual particles in the avalanche arrive at that position. 
This is clearly seen in Figs.\ \ref{spatio}(c) and \ref{spatio}(d) by the 
curved trajectories of particles near the front which are initially at rest.
To quantify these qualitative differences between particle and front motion,
the average velocity of individual particles $u_p$ near the avalanche front 
and the front velocity $u_f$ are measured for numerous avalanches. At a given
plane inclination $\theta$ the particle velocity as a function of the front 
velocity shows a linear dependence, as indicated in the inset of Fig.\ 
\ref{velopart-per-front}.
The ratio $u_p/u_f$ is plotted in Figs.\ \ref{velopart-per-front}(a) and
\ref{velopart-per-front}(b) as a function of the plane inclination for
sand/glass-beads and  copper particles, respectively.  For glass beads and sand
particles, the separation is very clean with the spherical particles moving 
slower than the front with $1 > u_p/u_f \approx 0.6$ whereas the irregular 
sand particles overtake the front with $1 < u_p/u_f \approx 1.3$.  There may 
be a slight downward trend with increasing $\theta$ for irregular particles 
but the data do not differentiate that trend from a near constant ratio.

The ratios $u_p/u_f$ are again less clearly separated for copper particles with
ratios close to one for the $\eta = 0.63$ and $\eta=0.5$ particles. 
For whatever reason - perhaps inter-particle friction - spherical copper particles 
are marginal with respect to the separation of particle velocity and front 
velocity.  The irregular copper particles have $u_p/u_f \approx 1.6$, 
considerably greater than one, and are well separated from the more 
spherical particles.  For the irregular particles there is a definite 
decrease in $u_p/u_f$ with increasing $\theta$.

\begin{figure}[ht]
\vspace*{0.1cm}
\resizebox{85mm}{!}{
\includegraphics{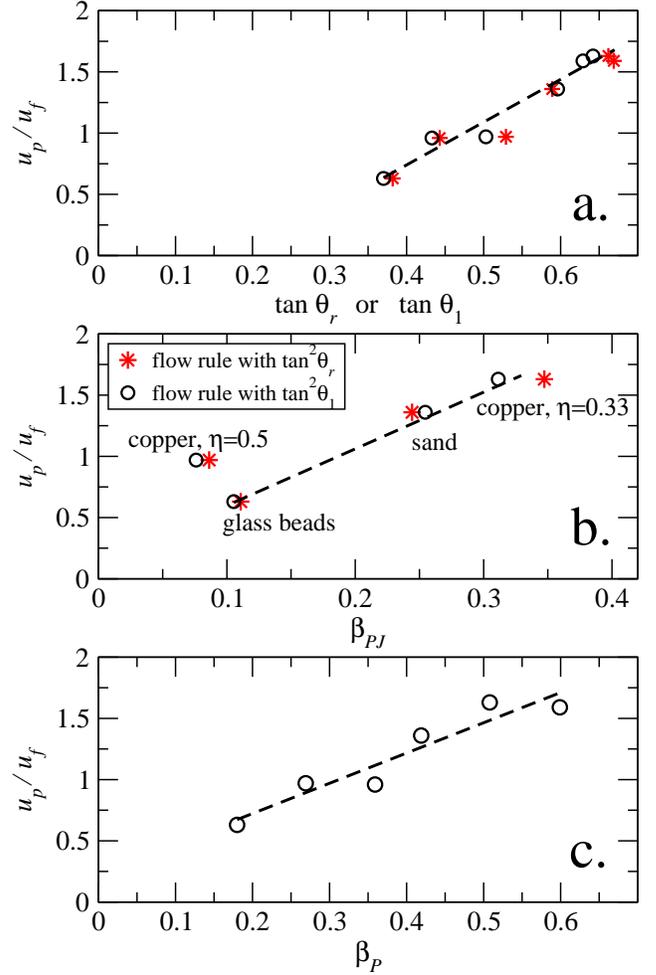}
}
\caption{(Color Online) Ratio of the particle velocity $u_p$ and front velocity $u_f$
as a function of (a) $\tan \theta_r$ (*) or $\tan \theta_1$ ($\circ$) or
(b) the slope $\beta_{PJ}$ of PJFR \cite{boec2007} or (c) the slope $\beta_{P}$ of PFR 
for sand, glass beads and copper with $\eta = 0.33$ and  $\eta = 0.5$.
In (b) the two data symbols correspond to the cases when $\beta_{PJ}$ was derived using
$\theta_r$ (*) or $\theta_1$ ($\circ$). 
}
\label{up-per-uf-flowruleslope}
\end{figure}

To quantify how the ratio $u_p/u_f$ varies with grain shape irregularity,
we again consider correlations between the ratio $u_p/u_f$ and the angle of 
repose $\theta_r$ (or $\theta_1$).  There is a linear increase of $u_p/u_f$ as 
a function of $\tan\theta_r$ (or $\tan\theta_1$) with a slope of about 3.5 as
shown in Fig.\ \ref{up-per-uf-flowruleslope}(a) for the 6 materials presented
in Fig.\ \ref{velopart-per-front}.
We also compare how the ratio $u_p/u_f$ varies with the equation relating $u$ 
and $h$ in the continuous flow regime, i.e., the granular flow rule.
Earlier, we reported values of the slope $\beta_{PJ}$ in Eq.\ \ref{PJFR}
for various materials \cite{boec2007} and found a systematic increase 
of $\beta_{PJ}$ with increasing grain shape irregularity, i.e., with increasing 
$\theta_r$ (or $\theta_1$).
Plotting $u_p/u_f$ as a function of $\beta_{PJ}$, we again find a systematic
linear increase with a slope of about 2.8. The copper sample with $\eta = 0.5$
seems to be anomalous, perhaps related to the peculiar dynamics of this set of 
copper in that it is the only copper sample emitting strong sound during shearing.
This emission is similar to but much stronger than the sand from the Kelso
dune which is known to be an example of ``booming sand dunes'' \cite{doma2006}.
Interestingly this peculiarity is not reflected in the dependence of $u_p/u_f$
on $\theta_r$, see Fig.\ \ref{up-per-uf-flowruleslope}(a).

Although the Pouliquen flow rule (Eq. \ref{PFR}) does not perfectly 
describe the flow properties of homogeneous flows, especially for the case of 
copper particles \cite{boec2007}, for the classification of avalanches we have 
determined $\beta_P$ by taking only data for relatively shallow flows where $h/h_s<10$.
Plotting $u_p/u_f$ as a function of $\beta_{P}$ 
(see Fig. \ref{up-per-uf-flowruleslope}(c)) we again find a systematic linear increase.
We will come back to this observation in Sec.\ \ref{theory}.

\begin{figure}[ht]
\resizebox{85mm}{!}{
\includegraphics{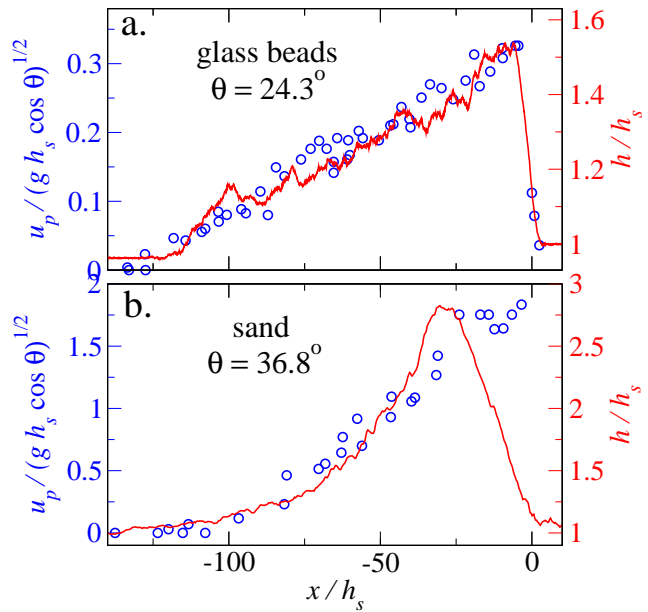}
}
\caption{(Color Online) Avalanche height profiles (continuous lines) taken with the help of a
laser sheet and velocity profiles (o) measured from space time plots for
(a) glass beads (for $\theta=24.3^\circ$ and $h_s=0.178$cm)
and (b) sand (for $\theta=36.8^\circ$ and $h_s=0.12$cm). 
}
\label{height-veloprofiles}
\end{figure}

Finally, we consider the detailed profiles of velocity and height along the 
propagating direction of the avalanche. The particle velocity $u_p(x)$ is obtained 
from space-time plots and the height profile from laser line measurements.  The 
measurements are not made simultaneously but can be compared for avalanches with 
similar properties.  We show profiles for glass beads and coarse sand in
Figs.\ \ref{height-veloprofiles}(a) and \ref{height-veloprofiles}(b) with 
scaled velocity and height on opposite axes.  The differences for sand and glass 
beads are striking. For glass beads, both velocity and height are approximately 
linear up to the maximum and then fall quickly over a steep but continuous front 
with a width of order 10$h_s$. The slope of the surface behind the front is only 
slightly shallower than the unperturbed layer with $\delta \theta/\theta \approx -0.007$ 
whereas the fractional angular increase near the front is about 0.1, i.e., a 
difference of about 3-4$^\circ$ relative to the plane inclination angle.

For sand avalanches, a quite different picture emerges as shown in 
Fig.\ \ref{height-veloprofiles}(b). First, the velocity maximum and the height
maximum occur at different values of downstream distance with height peaking 
before velocity.  Second, the height and velocity increase faster than linear 
from the back of the avalanche towards the front.  Although we have adjusted 
the axes to align the maximal velocity and height of the avalanche, one can 
still see that the functional dependence for scaled velocity greater than one 
and $h/h_s \ge 2$ are different.  In other words, velocity and height are 
proportional for small values near the back of the avalanche but separate above
values that, interestingly, differentiate in the mean between the spherical and
irregular avalanche behavior. Despite the quite different width of the sand front,
of order 25 $h_s$, the fractional angular increase at the front is only slightly 
shallower than for glass beads, i.e., $\delta \theta/\theta \approx 0.07$.  
The difference in profile for the sand arises from the nature of the particles 
near the front.  Whereas the glass beads form a compact avalanche with a sharp 
but distinct front, the sand particles overtake the front as defined by the 
motion of the layer underneath creating a dynamics similar to a breaking wave.
In the theory section below, we explore a depth-averaged approach for explaining
the different avalanche behavior of spherical and irregular granular particles.

\begin{figure}[ht]
\resizebox{85mm}{!}{
\includegraphics{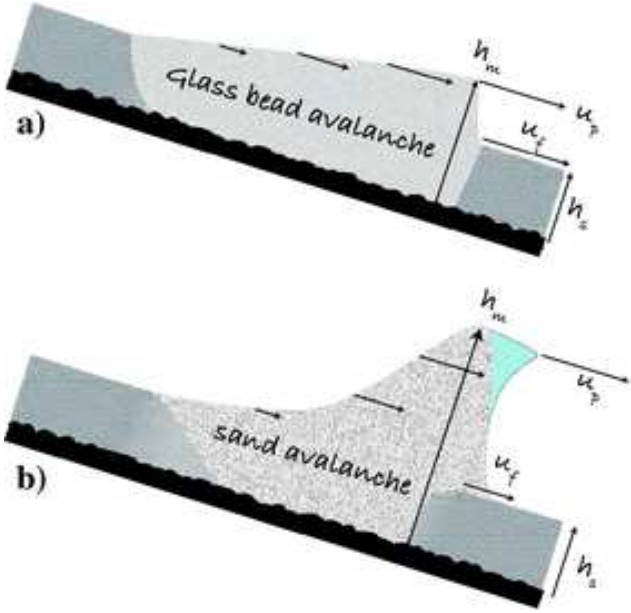}
}
\caption{(Color Online) Schematic view of avalanches formed in materials consisting
of (a) spherical particles (e.g., glass beads) and (b) irregular shaped particles 
(such as sand). The horizontal scale (along the slope) is compressed by about a 
factor of 30 relative to the scale perpendicular to the plane.  The tip region 
in b) represents low density material that spills over the front with speeds
greater than the front speed. Note the difference in the mobilization of the 
underlying layer, especially at the avalanche front.
}
\label{avalanche-schem}
\end{figure}

Although we cannot directly measure the interior profile of the avalanche, it is
useful to provide a schematic illustration of the different type of avalanches 
to summarize what we have learned about them.  In Fig.\ \ref{avalanche-schem}, 
we show the salient features of the two avalanche types. The case of shallower 
avalanches formed in materials with spherical particles is shown in 
Fig.\ \ref{avalanche-schem}(a). Here the granular layer fails, i.e., starts moving,
just ahead of the avalanche (of order 5-10 $h_s$), leading to slower maximal 
particle velocities than the front velocity. The velocity and height of the 
avalanche are proportional and the variation of both quantities is linear in the
down-plane direction.  The form of the avalanche is reminiscent of a viscous 
Burgers shock as discussed in more detail below.  In contrast, in 
Fig.\ \ref{avalanche-schem}(b) a sand avalanche is visualized having a larger 
maximal particle velocity than the front velocity and a considerably higher 
avalanche height.  The avalanche seems to behave like a ``breaking wave''
with low density particles spilling over the crest at speeds higher than the 
front speed.  The material directly under the front seems to be solid with the 
avalanche more slowly entraining the underlying material.  Direct measurements 
of this entrainment and the form of the avalanche profile would be very useful 
but are beyond the scope of the present work.

\section{Theory}
\label{theory}

The theory of avalanche behavior is complicated by the existence of a moving 
fluid phase in contact with a solid-like immobile granular phase.  Thus, a 
complete theoretical picture of avalanche behavior certainly requires a multi-phase 
approach.  On the other hand, since the granular material is moving over most of 
its extent except for small regions at the front and back, a simpler approach neglects
the solid state ahead and behind the avalanche. We can then use a depth-averaged 
approximation, leading to the Saint-Venant shallow flow equations, adapted for 
granular media by Savage and Hutter \cite{sahu1989}. This approach was used in the 
analysis of wave formation in dense granular flows \cite{fopo2003} and was applied  
by us to sand and glass-bead avalanches described previously \cite{boha2005}.  

For a flow of height $h$ and mean velocity $\overline u$,  granular flow down a
plane, with the plane parallel to the $x$-direction, is described by the
averaged equations for mass:

\begin{equation}
\frac{\partial h}{\partial t} + \frac{\partial (h\overline u)}{\partial x}=0
\label{saintvenant1}
\end{equation}

and momentum balance:

\begin{equation}
\frac{\partial (h\overline u)}{\partial t}+
\alpha \frac{\partial (h\overline u^2)}{\partial x} =
\left( \tan \theta -\mu (\overline u,h) -K \frac{\partial h}{\partial x} 
\right) gh \cos \theta.
\label{saintvenant2}
\end{equation}

The value of $\alpha$ is determined by the profile of the flow,
$\alpha=1$ for plug flow (as in Ref.\ \cite{sahu1989}),
$\alpha = 4/3$ for a linear flow profile, or 5/4 for a convex Bagnold 
profile \cite{doan1999}. The parameter $K$ is determined by the ratio of the
normal stresses in the flow: the stress parallel to the bed, $\sigma_{xx}$,
and that perpendicular to the bed, $\sigma_{zz}$.
Numerical results show that $K \equiv \sigma_{xx}/\sigma_{zz} \approx 1$
for steady-state flows \cite{sier2001}.  

The complicated part of this analysis concerns the friction coefficient 
$\mu(\overline u,h)$ \cite{fopo2002}.  For example, if the layer is too thin 
at a particular $\theta$, the flow stops and a friction coefficient appropriate 
for the flowing state is no longer valid. This transition from dynamic to static 
friction and the resulting yield condition for the solid phase in front of and 
behind the avalanche are not included in our approach and would be difficult 
to handle in a depth-averaged fashion.  We will consider the consequences of 
this assumption later. For now we assume that the friction coefficient 
$\mu(\overline u,h)$ is determined by the requirement that the steady flow obeys the 
rheology specified in Eq.~(\ref{Pouli1}), and will thus vary with the particle
type. For simplicity, we consider only the Pouliquen flow rule; the 
Pouliquen-Jenkins flow rule adds algebraic complexity without improved 
understanding of this model of avalanche behavior.

From the flow rule relation for steady-state flows (Eq. (\ref{PFR}))
we obtain an expression for $h_s$ in terms of $\overline u$ and $h$:

\begin{equation}
h_s=\frac{\beta_P h^{3/2}\sqrt{g}}{\overline u+\gamma\sqrt{gh}} \ .
\label{Pouli1}
\end{equation}

The avalanches we will describe are observed for angles not very far from
$\theta_1$ (i.e., $\tan \theta / \tan \theta_1<1.4$).  In that range, the
relationship
\begin{equation}
h_s = \frac{a d}{\tan \theta - \tan \theta_1}
\label{hs2}
\end{equation}
works very well in fitting the data for all the materials. The values for the
two fitting parameters $a$ and $\theta_1$ are shown in Table \ref{parameters} for
sand, glass, and copper.  As $\mu = \tan\theta$ we obtain from (\ref{hs2})

\begin{equation}
\mu = \tan\theta = \tan\theta_1 + \frac{a d}{h_s} \ .
\label{mu1}
\end{equation}

The slower avalanches we describe correspond to the case of low Froude numbers 
${\Froude} = \overline u/\sqrt{gh\cos\theta}$.  For example,  glass beads avalanches 
correspond to $\Froude \approx 0.4$.  Since the left hand side (LHS) of 
Eq.\ (\ref{saintvenant2}) scales like $\Froude^2 \approx 0.16$, we can set the
LHS to zero.  Plugging the expression for $\mu$ (\ref{mu1}) into the right hand side (RHS), we get:

\begin{equation}
\tan\theta-\tan\theta_1- \frac{\alpha d}{h_s}-K\frac{\partial 
h}{\partial x} = 0 \ .
\label{eq2deriv}
\end{equation}

Substituting the expression for $h_s$ (\ref{Pouli1}) into 
(\ref{eq2deriv})  gives an expression for $\overline u$:

\begin{equation}
\overline u = \frac{\sqrt{g}}{\alpha d} \left(\tan\theta-\tan\theta_1 - 
K\frac{\partial h}{\partial x}\right) \cdot \beta_P h^{3/2}
- \gamma \sqrt{gh} \ .
\label{u}
\end{equation}

Finally we use this form for $\overline u$ to substitute into (\ref{saintvenant1}):

\begin{equation}
\frac{\partial h}{\partial t} +
\frac{\partial }{\partial x}\left[\frac{\sqrt{g}}{\alpha d}
\left(\tan\theta-\tan\theta_1 -
K\frac{\partial h}{\partial x}\right) \cdot \beta_P h^{5/2}
- \gamma h\sqrt{gh} \right]
\nonumber
\end{equation}

\begin{equation}
=0
\label{result1}
\end{equation}

which yields:

\begin{equation}
\frac{\partial h}{\partial t} +
  a(h) \cdot \frac{\partial h}{\partial x} =  \frac{K\beta_P \sqrt{g}}{\alpha d}
\frac{\partial}{\partial x} \left[h^{5/2} \left(\
\frac{\partial h}{\partial x} \right) \right]
\nonumber
\end{equation}

where
\begin{equation}
\ \ \ \ a(h) = \sqrt{gh} \left(\frac{\frac{5}{2}\beta_P h }{h_s} -
\frac{3}{2}\gamma \right).
\label{result5}
\end{equation}

This equation has solutions similar to those of Burger's equation 
\cite{wh1999}. Thus, there is a solution consisting of a single hump propagating 
down the slope with velocity $a(h)$, with a smooth structure determined by the 
competition between this nonlinear velocity term on the LHS and the dissipative 
term (RHS) of Eq.~(\ref{result5}).  Numerical solution of this equation yields a 
fairly linear ramp behind the front, very similar to Fig.\ \ref{height-veloprofiles}(a) 
for glass beads.  Further, one can estimate the width of the front by balancing 
the dispersive term and the dissipative term, taking $\partial h/\partial x \sim h/\ell$ and 
$\partial^2 h/\partial x^2 \sim h/\ell^2$ where $\ell$ is the front width.  
Inserting values for the parameters in Eq.\ (\ref{result5}) yields a scaled front 
width $\ell/h_s \approx 6$, consistent with the height profile for glass 
beads in Fig.\ \ref{height-veloprofiles}(a).  We can also compare the 
predicted group velocity $a(h)$ with the measured avalanche front
velocity for glass beads.  Using the peak height $h_m$ to evaluate $a(h)$,
we obtain $u_f/a(h) \approx 0.6$, again consistent agreement between theory
and experiment. 

The full equations of mass and momentum conservation are
hyperbolic with characteristic velocities

\begin{equation}
c_{\pm}=\overline u\left(\alpha \pm \sqrt{\alpha (\alpha-1) +\frac{K}{\Froude^2}}\right).
\label{resultc}
\end{equation}

\noindent If the velocity appearing in Eq.\ (\ref{result5}) does not
obey  $a < c_+$, then Eq.~(\ref{result5}) predicts a structure that
moves faster than the maximum rate at which information can be
propagated in the full system of equations, which is clearly
impossible. For faster avalanches such as sand, the small Froude
number limit is not valid since Fr $>$ 1. In these circumstances, the 
Burger's type solution transforms itself into a truly discontinuous 
solution traveling at velocity $c_+$ \cite{wh1999}, which is described 
by the full system Eqs.\ (\ref{saintvenant1}-\ref{saintvenant2}) rather than by 
Eq.\ (\ref{result5}). 

\begin{figure}[ht]
\resizebox{85mm}{!}{
\includegraphics{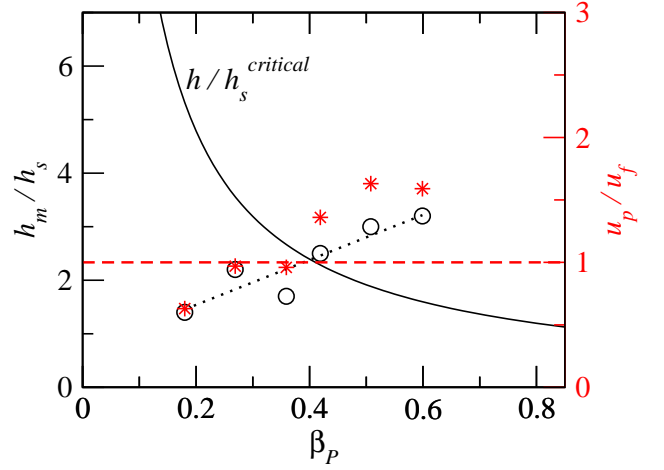}
}
\caption{(Color Online) Dimensionless avalanche height $h_m/h_s$ ($\circ$) and the ratio 
of the particle and front velocities $u_p/u_f$ (*) as a function of $\beta_P$ for sand, 
glass beads and the four copper samples. The continuous line corresponds to the criterion
$a = c_+$, and the dotted line is a linear fit to the $h_m/h_s(\beta_P)$ data. A horizontal
dashed line denotes the value of $u_p/u_f=1$.
}
\label{hperhs-beta}
\end{figure}

We can directly test the conditions for which  $a < c_+$ is valid
in terms of avalanche height. The normalized avalanche height $h_m/h_s$ as taken from 
the slope of the fits in Figs.\ \ref{height-layerthickness}(a) and 
\ref{height-layerthickness}(b) is shown in Fig.\ \ref{hperhs-beta} for glass beads, 
sand and all four copper samples as a function of the flow rule slope $\beta_P$ 
(using $\gamma=0$). We find a systematic increase of $h_m/h_s$ as a function of $\beta_P$, 
as we did when plotting $h_m/h_s$ as a function of $\tan\theta_r$ 
(see Fig.\ \ref{height-layerthickness}(c)).
On the other hand, the boundary between the two regimes corresponding to
$a = c_+$ results in a critical thickness $h/h_s^{critical}$ of the flow that 
strongly decreases with increasing $\beta_P$ (see the continuous line in 
Fig. \ref{hperhs-beta}).  The range below $h/h_s^{critical}$ corresponds to 
smooth solutions of Eq.\ (\ref{result5}), with $a < c_+$, whereas above 
$h/h_s^{critical}$ theory predicts discontinuous solutions of the full system
traveling at velocity $c_+$.  The line corresponding to the normalized avalanche
height $h_m/h_s$ as a function of $\beta_P$ crosses this boundary, 
and the data points for sand and the two anisotropic (dendritic) copper grains 
fall above the curve, whereas data points for glass beads and the spherical copper 
samples fall below the boundary.
This observation matches well with the result that the ratio of the particle 
velocity and the front velocity of the avalanche $u_p/u_f$ also increases 
with increasing $\beta_P$ and crosses the value 1 at about the same location 
corresponding to $\beta_P\approx0.4$ (see Fig. \ref{hperhs-beta}).

Thus, we conclude that the glass bead avalanches, or more generally avalanches 
formed in materials with small $\beta_P$, reflect smooth solutions of
Eq.~(\ref{result5}), with $a < c_+$, whereas avalanches in sand or copper samples 
with highly anisotropic grains, or more generally in materials with higher $\beta_P$,
represent discontinuous solutions of the full system traveling at velocity $c_+$.
The latter avalanches propagate into a quiescent bed because they are traveling at
the characteristic velocity for the medium.  In the experiments the signature
of these ``discontinuous'' solutions is that grains move faster than the front
speed, spilling or breaking ahead of the front in a low density ``foam-like'' 
phase. The crossover between these two regimes happens at $\beta_P\approx 0.4$.
The glass bead avalanches are analogous to ``flood waves" in river flows,
whereas the sand avalanches are analogous to ``roll waves" in these
flows \cite{wh1999,dr1949}.
Note that ahead of the flowing avalanche, the moving material propagates
into a material at rest, which is presumably in a state close to the
critical Mohr-Coulomb state \cite{nedderman}. Unlike the flowing state,
for which $\sigma_{xx} \approx \sigma_{zz}$, in this critical state
$\sigma_{xx} > \sigma_{zz}$. Thus the transition region, in which
the flow accelerates from rest into a pseudo-steady state described
by the continuum theory, can be viewed as a region of passive Rankine
failure, through which the compressive stress parallel to the bed,
$\sigma_{xx}$, is decreasing with time. The mechanics of this region
are complex, and cannot be described by the Saint-Venant equations alone.

We should point out that in the linear theory of the instability of
steady flows, developed by Forterre and Pouliquen, the criterion
$a < c_+$ corresponds to the stable regime of these flows with
respect to wave disturbances \cite{fopo2003}. Thus, our observation
of discontinuous avalanches for sand and smooth avalanches for
glass beads dovetails nicely with their observation that steady
flows of sand are far more unstable to such disturbances than are
steady flows of glass beads.

The theory as presented has several nice features that semi-quantitatively 
describe the dynamics of avalanches in a range of materials with varying 
shapes and inter-particle friction.  Nevertheless, the theory fails to account
for several important aspects of the flow.  First, the generalized Burgers 
equation fails to capture the soliton-like nature of the individual avalanches,
namely that they travel at constant speed and maintain uniform shape.  One can 
include more physics by following Forterre and Pouliquen \cite{fopo2002} in 
providing a parameterization of the friction coefficient that accounts for 
the static friction component and hysteresis between flowing and solid phases.
Numerical simulations of either the full 2D equations or a 1D version of the
Saint-Venant equation with a more realistic friction, fail to resolve this 
difference \cite{SlimPC}.  In particular, for glass beads one obtains
a solitary localized solution but one with a flat rather than linear height 
profile behind the front.  In that case, the speed is controlled by a front 
condition rather than by the mechanism consistent with our theory. 
We find empirically that the predicted velocity $a(h)$ is greater than the 
measured $u_f$, so perhaps a similar control is at play for the glass beads.
For sand avalanches, full numerical simulations yield a form similar to our
theory but again the height decays rather than assuming a constant profile. 
Thus, although the theory presented here captures some of the ingredients of 
our experimental avalanches, the two phase nature of the avalanches is probably 
necessary for a more quantitative description of our results.

\section{Conclusions}
\label{conclusion}

The main conclusion of our experimental study of  avalanches is that the form 
of the avalanche depends on the granular particles. Differences are found not only
at the quantitative level, but in the qualitative features (front propagation, 
avalanche shape, grain dynamics) which reflect the dramatic differences between 
smooth grains such as glass beads and irregular grains like sand. Measuring 
the basic avalanche properties such as velocity and size, we find a systematic 
linear increase of the normalized avalanche peak height $h_m/h_s$ with increasing 
$\tan \theta_r$ or $\beta_P$, where $\theta_r$ is the angle of repose of the 
material and $\beta_P$ is the slope of the depth-averaged flow rule, both measures
of grain shape irregularity.  Similarly, the slope of the dimensionless velocity 
versus size curves increases linearly with increasing $\tan \theta_r$, suggesting 
that the avalanche velocity increases with increasing avalanche size systematically
faster for more anisotropic grains.
These two relations enable us to predict typical avalanche sizes and the 
growth rate of avalanche velocity as a function of  increasing size for various materials, 
by simply measuring $\theta_r$ or $\beta_P$.

When focusing on the dynamics of individual grains (and comparing it to the 
propagation of the avalanche), another characteristic difference is found for 
spherical and non-spherical particles.  Particles with irregular shape move 
considerably faster than the avalanche front, indicating that inertia is an 
important ingredient in these faster moving avalanches.  For avalanches formed 
by spherical particles, on the other hand, front propagation is transmitted 
through the contact points (force chains) near the front, thereby inducing 
motion through a collective yield (failure) condition at the front.  In the 
latter case,  the velocity of individual grains is considerably smaller than 
the front velocity. When quantifying this important property of avalanches we 
again find a direct correspondence between $u_p/u_f$ and either $\theta_r$, the angle
of repose of the material, or the slope of the depth averaged flow rule $\beta_P$
(or $\beta_{PJ}$). The value $u_p/u_f$ increases linearly with increasing 
$\tan \theta_r$ or $\beta_P$.  These relationships again provide a tool for 
predicting the dynamical properties of avalanches by simply measuring 
$\theta_r$ or $\beta_P$ for a given material. The direct correspondence
should be valid for moderately polydisperse materials, but may not be valid 
for highly polydisperse materials where segregation effects become important.

In Section \ref{theory} we have shown that a simple depth-averaged description 
can capture a strong change in the dynamical behavior of avalanches if we know 
the basic (depth-averaged) rheology of the material.  Also the striking 
difference between the behavior of avalanches formed by spherical and nonspherical
particles warns us that when modeling granular avalanches or granular dynamics 
in general, predictions obtained by the simplest models neglecting shape irregularities
might not be valid for systems with non-spherical particles.  Our measurements, 
however, provide numerous additional quantitative features of avalanches for grains
with different rheologies that cannot be captured by the simple theory presented here,
leaving room for further theoretical or numerical studies that can provide a better
understanding of avalanching behavior.

The authors benefited from discussions with Brent Daniel, Anja Slim, Michael Rivera
and Michael Stepanov.  One of us (R.E.E.) acknowledges the support of the Aspen
Center for Physics.  This research was funded by the U.S. Department of Energy under
Contracts No.\ W-7405-ENG and No.\ DE-AC52-06NA25396.  T.B. acknowledges support 
by the Bolyai J\'anos research program and the Hungarian
Scientific Research Fund (Contract No.\ OTKA-F-060157).


\begin{thebibliography}{99}
\bibitem{jana1996} H.M. Jaeger, S.R. Nagel  and R.P. Behringer,
Rev. Mod. Phys. {\bf 68} 1259 (1996).

\bibitem{fe1995}
J. Feder, Fractals {\bf 3}, 431 (1995).

\bibitem{ra1990}
J. Rajchenbach, Phys. Rev. Lett. {\bf 65}, 2221 (1990).

\bibitem{ra2002}
J. Rajchenbach, Phys. Rev. Lett. {\bf 88}, 014301 (2001).

\bibitem{cogo2005}
S. Courrech du Pont, R. Fischer, P. Gondret, B. Perrin and M. Rabaud,
Phys. Rev. Lett. {\bf 94}, 048003 (2005).

\bibitem{boda2002} D. Bonamy, F. Daviaud and L. Laurent, Phys. of 
Fluids {\bf 14}, 1666 (2002).

\bibitem{du2000} D. J. Durian, J. Phys. Cond. Mat. {\bf 12}, A507 (2000).

\bibitem{ledu2000}
P.-A. Lemieux and D. J. Durian, Phys. Rev. Lett. {\bf 85}, 4273 (2000).

\bibitem{neag2003} N. Nerone, M.A. Aguirre, A. Calvo, D. Bideau and 
I. Ippolito,
Phys. Rev. E. {\bf 67}, 011302 (2003).

\bibitem{liha1999} S.J. Linz, W. Hager and P. H\"anggi, Chaos {\bf 
9}, 649 (1999).

\bibitem{si2005} L.E. Silbert, Phys. Rev. Lett. {\bf 94}, 098002 (2005).

\bibitem{koin2001} T.S. Komatsu, S. Inagaki, N. Nakagawa and S. Nasuno,
Phys. Rev. Lett. {\bf 86}, 1757 (2001).

\bibitem{ando2001} B. Andreotti and S. Douady, Phys. Rev. E. {\bf 
63}, 031305 (2001).

\bibitem{arra1999} A. Aradian, E. Rapha\"el and P.G. de Gennes, Phys. 
Rev. E. {\bf 60}, 2009 (1999).

\bibitem{emcl2005}
T. Emig, P. Claudin and J.P. Bouchaud, Phys. Rev. E {\bf 71}, 031305 (2005).

\bibitem{ivre2000} R.M. Iverson, M.E. Reid, N.R. Iverson, R.G. 
LaHusen, M. Logan, J.E. Mann
and D.L. Brien, Science {\bf 290}, 513 (2000).

\bibitem{dado1999}
A. Daerr and S. Douady, Nature {\bf 399}, 241 (1999).

\bibitem{da2001}
A. Daerr, Phys. of Fluids {\bf 13}, No.7, 2115 (2001).

\bibitem{da2000}
A. Daerr,  Ph.D. thesis, University of Paris VII (France), (2000).

\bibitem{arts2001}
I.S. Aranson and L.S. Tsimring, Phys. Rev. E {\bf 64}, 020301(R) (2001).

\bibitem{ra2002aug}
J. Rajchenbach, Phys. Rev. Lett. {\bf 89}, 074301 (2002).

\bibitem{feth2004} G. F\'elix and N. Thomas,
Earth and Planetary Science Letters {\bf 221}, 197 (2004).

\bibitem{mala2006}
F. Malloggi, J. Lanuza, B. Andreotti and E. Cl\'ement,
Europhys. Lett. {\bf 75}, 825 (2006).

\bibitem{clma2007}
E. Cl\'ement, F. Malloggi, B. Andreotti and I.S. Aranson,
Granular Matt. {\bf 10}, 3 (2007).


\bibitem{arma2006}
I.S. Aranson, F. Malloggi and E. Cl\'ement,
Phys. Rev. E {\bf 73}, 050302(R) (2006).

\bibitem{po1999}
O. Pouliquen, Phys. of Fluids {\bf 11}, No.3, 542 (1999).

\bibitem{pona1996}
O. Pouliquen and N. Reanut, J. Phys. II. France {\bf 6}, 923 (1996).

\bibitem{ba1954}
R.A. Bagnold, Proc. R. Soc. London, Ser. A {\bf 255}, 49 (1954).

\bibitem{gdrmidi2004} GDR MiDi, Eur. Phys. J. E {\bf 14}, 341 (2004).

\bibitem{sier2001} L.E. Silbert, D. Ertas, G.S. Grest, T.C. Halsey, D. Levine
and S.J. Plimpton,  Phys. Rev. E {\bf 64}, 051302-1 (2001).

\bibitem{sila2003} L.E. Silbert, J.W. Landry and G.S. Grest,
Phys. of Fluids {\bf 15}, 1 (2003).

\bibitem{boec2007}
T. B\"orzs\"onyi and R.E. Ecke, Phys. Rev. E {\bf 76}, 031301 (2007).

\bibitem{fopo2003}  Y. Forterre and O. Pouliquen, J. Fluid Mech. {\bf 
486}, 21 (2003).

\bibitem{je2006}
J.T. Jenkins, Phys. of Fluids {\bf 18}, 103307 (2006).

\bibitem{anda2002} B. Andreotti, A. Daerr and S. Douady, Phys. of 
Fluids {\bf 14}, 415 (2002).

\bibitem{babe1989} G.W. Baxter, R.P. Behringer, T. Fagert and G.A. Johnson,
Phys. Rev. Lett. {\bf 62}, 2825 (1989).

\bibitem{cl2008} P.W. Cleary, Powder Technology {\bf 179}, 144 (2008).

\bibitem{boha2005}
T. B\"orzs\"onyi, T.C. Halsey and R.E. Ecke, Phys. Rev. Lett. {\bf 
94}, 208001 (2005).

\bibitem{boec2006}
T. B\"orzs\"onyi and R.E. Ecke, Phys. Rev. E {\bf 74}, 061301 (2006).

\bibitem{movieweb}
Movies of avalanches taken with a high speed camera can be downloaded from:
http://www.szfki.hu/btamas/gran/avalanche.html or
http://www.lanl.gov/orgs/mst/MST10/fluid $\underline{\ 
}$dynamics/granular.shtml

\bibitem{erha2002} D. Ertas, and T.C. Halsey, Europhys. Letts. {\bf 
60}, 931 (2002).

\bibitem{doma2006}
S. Douady, A. Manning, P. Hersen, H. Elbelrhiti, S. Protiere,
A. Daerr, B. Kabbachi, Phys. Rev. Lett. {\bf 97}, 018002 (2006).

\bibitem{sahu1989}  S.B. Savage  and K. Hutter, J. Fluid Mech. {\bf 
199}, 177 (1989).

\bibitem{doan1999} S. Douady, B. Andreotti and A Daerr, Eur. Phys. J. 
B.  {\bf 11}, 131 (1999).

\bibitem{fopo2002}  O. Pouliquen and Y. Forterre, J. Fluid Mech. {\bf
453}, 133 (2002).

\bibitem{wh1999}  G.B. Whitham, {\it Linear and Nonlinear Waves}
(Wiley, New York, 1999).

\bibitem{dr1949}   R.F. Dressler,  Comm. Pure Appl. Math.{\bf 2}, 149 (1949).

\bibitem{nedderman} R.M. Nedderman, {\it Statics and Kinematics of
Granular Materials} (Cambridge Univ. Press, 1992, New York) Chapter 2.

\bibitem{SlimPC}  A. Slim, Private Communication.

\end{thebibliography}
\end{document}